\documentclass[12pt]{article}
\usepackage{color,amssymb,epsfig,verbatim}
\usepackage{graphicx} 
\usepackage{bm}
\usepackage{multirow}
\usepackage{amsmath}
\usepackage{float}
\usepackage{algorithm}
\usepackage{algpseudocode}
\usepackage{lscape}
\usepackage[table,xcdraw]{xcolor}
\usepackage{natbib}
\usepackage{hyperref}
\usepackage{accents}
\newtheorem{proposition*}{Proposition}
\hypersetup{colorlinks,linkcolor={black},citecolor={gray},urlcolor={red}}

\renewcommand{\baselinestretch}{2}  

\definecolor{jade}{RGB}{0,168,107}
\newcommand{\qz}[2][]{\textcolor{jade} {[#2] {\bf \scriptsize (QZ\ifx\relax#1\relax\else-#1\fi)}}}

\definecolor{mlblue}{RGB}{0,114,178}
\newcommand{\ml}[2][]{\textcolor{mlblue}{[#2] \textbf{\scriptsize (ML\ifx\relax#1\relax\else-#1\fi)}}}

\definecolor{kfyellow}{RGB}{180,130,0}
\newcommand{\kf}[2][]{\textcolor{kfyellow}{[#2] \textbf{\scriptsize (KF\ifx\relax#1\relax\else-#1\fi)}}}

\def\log{\hbox{log}}

\def\dfrac#1#2{{\displaystyle{#1\over#2}}}

\def\boxit#1{\vbox{\hrule\hbox{\vrule\kern6pt
          \vbox{\kern6pt#1\kern6pt}\kern6pt\vrule}\hrule}}

\def\bse{\begin{eqnarray*}}
\def\ese{\end{eqnarray*}}
\def\be{\begin{eqnarray}}
\def\ee{\end{eqnarray}}
\def\bq{\begin{equation}}
\def\eq{\end{equation}}
\def\bse{\begin{eqnarray*}}
\def\ese{\end{eqnarray*}}

\newtheorem{Assum}{\underline{\bf Assumption}}

\newtheorem{proposition}{Proposition}
\newtheorem{Lem}{\underline{\bf Lemma}}
\newtheorem{Cor}{\underline{\bf Corollary}}

\newenvironment{proof}{\par\noindent{\bf Proof.}\ }{\hfill$\square$\par}

\DeclareMathOperator*{\argmax}{arg\,max}

\newcommand{\E}{\mathrm{E}}

\newcommand{\ind}{\mathbb{I}}
\newcommand{\normal}{\mathcal{N}}

\usepackage[para]{threeparttable}
\usepackage{rotating}

\begin{document}
\thispagestyle{empty}

\hfill\today \\ \\

\baselineskip=28pt
\begin{center}
{\LARGE{\bf Model-Based Co-Clustering for Censored Data}}
\end{center}
\baselineskip=14pt
\vskip 2mm
\begin{center}
Kejun Fang\footnote{\baselineskip=10pt Department of Mathematics and Statistics, McGill University, Montreal, Quebec, Canada}, Mincen Liu\footnote{\baselineskip=10pt Department of Epidemiology, Biostatistics and Occupational Health, McGill University, Montreal, Quebec, Canada H3A 1G1}, and Qihuang Zhang\footnotemark[2]\textsuperscript{*}

\end{center}

\begingroup
\renewcommand{\thefootnote}{*}
\footnotetext{\baselineskip=10pt Corresponding author: qihuang.zhang@mcgill.ca}
\endgroup
\bigskip
\vspace{8mm}

\begin{center}
{\Large{\bf Abstract}}
\end{center}
\baselineskip=17pt

Co-clustering jointly partitions the rows and columns of a data matrix. Standard co-clustering methods for continuous data typically assume that entries are observed exactly. Matrices with censored entries, such as assay measurements below a detection limit, are therefore commonly completed by substitution or imputation before co-clustering. Since the completed values are treated as exact, the estimated block means and partitions depend on the completion rule. We propose the censored Gaussian latent block model (CGLBM), in which each censored entry contributes the probability of its reported interval. We establish identifiability under a known Type I censoring design. Approximate standard errors for the block means follow from the Fisher information under censoring. CGLBM is fitted by a generalized expectation--maximization algorithm, and its row and column class numbers are selected by a BIC approximation to the integrated classification likelihood. In simulation studies, boundary substitution and truncated normal imputation attenuate the contrasts between block means, whereas the partition recovery of multiple imputation drops the most under heavy censoring. CGLBM preserves the contrasts, and its partition recovery drops the least. The method is implemented in the R package \texttt{CGLBM}.

\vspace{8mm}

\par\vfill\noindent

\par\medskip\noindent
\underline{\bf Short title}: CGLBM

\clearpage\pagebreak\newpage
\pagenumbering{arabic}

\newlength{\gnat}
\setlength{\gnat}{22pt}
\baselineskip=\gnat

\clearpage



\section{Introduction}\label{sec:introduction}

Studies that measure many analytes in each sample, such as multiplex immunoassays and biomonitoring surveys, produce a data matrix with samples as rows and analytes as columns. In such settings, a pattern shared by a group of rows may occur only across a group of columns, so clustering either dimension separately can miss the interactions that define the underlying structure. Co-clustering addresses this problem by partitioning the rows and the columns jointly, so that each pair of a row class and a column class defines a block of cells with a common pattern \citep{govaert2003,biernacki2023}. Existing methods recover the two partitions through spectral \citep{dhillon2001}, information-theoretic \citep{dhillon2003}, or matrix factorization \citep{ding2006} criteria. They have been used to group tumor samples together with marker genes \citep{kluger2003} and survey respondents together with questionnaire items \citep{selosse2019}. Biclustering methods, by contrast, search for submatrices that may overlap and need not cover the matrix \citep{cheng2000,lazzeroni2002,madeira2004}. The latent block model (LBM) assigns a probability distribution to each block and estimates the row and column memberships together with the block parameters \citep{govaert2003,biernacki2023}.

Many biomedical and environmental measurements are subject to lower or upper
detection limits. Censoring occurs when the exact measurement is not observed
but is known to lie below or above a known reporting limit. A data matrix may
therefore contain exact values together with censored reports. Standard co-clustering methods require a complete numerical matrix. A common response is to complete the matrix before co-clustering. Simple procedures replace censored reports by the corresponding detection limits or fixed fractions of those limits \citep{helsel2006,lubin2004}. Reports censored at the same limit then receive the same value even when their underlying measurements differ. This can distort means, variances, correlations, and group comparisons \citep{helsel2006,lubin2004}. For example, substituting the limit raises the estimated mean of a block whose measurements fall mostly below it, which shrinks its contrast with the other blocks, whereas substituting a fraction of the limit can move the estimate in either direction. When reporting limits vary, fixed substitution can also introduce artificial patterns or obscure existing patterns in the data \citep{helsel2006}. More elaborate completion methods draw plausible values from distributions restricted to the reported intervals, either to form one completed matrix or, through multiple imputation, to form several completed matrices \citep{lubin2004,faucheux2021}. These methods use the interval information, but they estimate the completed values separately from the row and column groups. As a result, the imputed values are drawn without regard to the partitions they are then used to estimate. Once the matrix has been completed, the imputed entries are treated as exact observations during co-clustering. The estimated block means and partitions therefore depend on the completion rule.

Compared to completing the data first, censored observations can be handled directly through likelihood-based models. Exact values contribute to density terms, whereas censored reports contribute probabilities over their reported regions. \citet{lee2012} developed expectation--maximization (EM) algorithms for Gaussian mixture models with truncated and censored data. Later work allowed heavy tails \citep{lachos2017}, mild outliers \citep{lin2026}, and matrix-valued observations \citep{tomarchio2026}. These mixtures cluster the observations, not the rows and the columns of one data matrix. Existing extensions have incorporated missing data into the LBM. \citet{selosse2020} assumed that the missing process is missing at random and imputed missing values within a stochastic EM (SEM)-Gibbs algorithm. \citet{corneli2020} developed a latent block model for ordinal categorical data while allowing the missingness process to be missing not at random (MNAR). \citet{frisch2022} combined the LBM with an MNAR missingness model. Missing entries and censored reports provide different information. A missing entry provides no numerical constraint on the unobserved value, whereas a censored report specifies known bounds on the underlying measurement. To our knowledge, no latent block model for continuous data uses these bounds.

We introduce the censored Gaussian latent block model (CGLBM) for co-clustering matrices containing exact and censored reports. CGLBM assigns a Gaussian distribution to each block defined by a row class and a column class. The cell likelihood uses a Gaussian density for an exact report and the Gaussian probability of the reported interval for a censored report. The reporting limits are fixed and known, and the row memberships, column memberships, and block parameters are estimated jointly without first completing the matrix. First, we establish identifiability of the Gaussian block parameters and the corresponding row and column classes under a known Type I censoring design, up to separate permutations of the row and column labels. Second, we develop a common free energy objective and optimize it using a unified generalized EM (GEM) procedure for hard and soft memberships. Hard memberships assign each row or column to a single class, whereas soft memberships assign weights across the possible classes. Third, we select the row and column class numbers through a Bayesian information criterion (BIC) approximation to the integrated classification likelihood. Fourth, we derive the Fisher information under censoring and show that every finite censoring threshold retains information about both the Gaussian mean and variance. We use this information to obtain approximate standard errors for fitted block means. In simulation studies, boundary substitution and truncated normal imputation attenuated the contrasts between block means, and the partition recovery of multiple imputation declined the most under heavy censoring. CGLBM preserved the contrasts, and its partition recovery declined the least. The method is implemented in the R package \texttt{CGLBM}.

The rest of the article is organized as follows: Section~\ref{sec:censored_gaussian_lbm} introduces CGLBM and studies its identifiability. Subsequently, Section~\ref{sec:estimation_fixed_candidate_model} develops the estimation procedure and the criterion for selecting the numbers of row and column classes. It also derives the Fisher information under censoring and the resulting approximate standard errors for fitted block means. The methodology is then assessed through simulation studies in Section~\ref{sec:simulation_studies}. Finally, Section~\ref{sec:discussion} summarizes the main results and provides direction for future research.



\section{Censored Gaussian latent block model}\label{sec:censored_gaussian_lbm}

\subsection{Latent row and column classes}\label{subsec:latent_row_column_classes}

Let $\mathbf{Y}=(Y_{ij})_{1\leq i\leq n,\,1\leq j\leq m}\in\mathbb R^{n\times m}$ denote the underlying continuous measurement matrix, where $n$ and $m$ are the numbers of rows and columns, respectively. Let $K$ and $L$ be fixed numbers of row and column classes. For $i=1,\ldots,n$, let $Z_i\in\{1,\ldots,K\}$ denote the class label of row $i$. For $j=1,\ldots,m$, let $W_j\in\{1,\ldots,L\}$ denote the class label of column $j$. Write $\mathbf Z=(Z_1,\ldots,Z_n)^\intercal$ and $\mathbf W=(W_1,\ldots,W_m)^\intercal$.

The Gaussian latent block model specifies the joint distribution of $\mathbf Y$, $\mathbf Z$, and $\mathbf W$ through the following assumptions:
\begin{enumerate}
\item 
The labels $Z_1,\ldots,Z_n,W_1,\ldots,W_m$ are mutually independent, with $\Pr(Z_i=k)=\pi_k$ and $\Pr(W_j=\ell)=\rho_\ell$.
Let
$\boldsymbol\pi=(\pi_1,\ldots,\pi_K)^\intercal$ and
$\boldsymbol\rho=(\rho_1,\ldots,\rho_L)^\intercal$
denote the row and column class proportion vectors, respectively.
\item Conditional on $\mathbf Z$ and $\mathbf W$, the cells $Y_{ij}$ are independent, and the distribution of $Y_{ij}$ depends only on its row label $Z_i$
and column label $W_j$.
\end{enumerate}

For $k=1,\ldots,K$ and $\ell=1,\ldots,L$, let
$\mu_{k\ell}$ and $\sigma^2_{k\ell}$ denote the mean and variance,
respectively, of the conditional distribution of $Y_{ij}$ given
$Z_i=k$ and $W_j=\ell$. Let
$\boldsymbol\mu=\{\mu_{k\ell}\}_{k,\ell}$ and
$\boldsymbol\sigma^2=\{\sigma^2_{k\ell}\}_{k,\ell}$ collect these means and
variances, respectively, and let
$\boldsymbol\theta=(\boldsymbol\pi,\boldsymbol\rho,\boldsymbol\mu,\boldsymbol\sigma^2)$ collect the model parameters.

Let $\mathbf z=(z_1,\ldots,z_n)^\intercal$ and $\mathbf w=(w_1,\ldots,w_m)^\intercal$ denote realizations of $\mathbf Z$ and $\mathbf W$. Then, the Gaussian latent block model \citep{govaert2003,lomet2018} is postulated by modeling the joint distribution of  $Z$ and $W$,
\begin{equation}\label{eq:label_prior}
p_{\boldsymbol\theta}(\mathbf z,\mathbf w)
:=
\prod_{i=1}^n\pi_{z_i}
\prod_{j=1}^m\rho_{w_j},
\end{equation}
and the conditional distribution of $Y_{ij}$ given by $Z$ and $W$ as a normal distribution. Namely,
\begin{equation}\label{eq:block_gaussian_law_revised}
Y_{ij}\mid Z_i=k,W_j=\ell
\sim
\normal(\mu_{k\ell},\sigma_{k\ell}^2),
\qquad
k=1,\ldots,K,
\quad
\ell=1,\ldots,L.
\end{equation}
The pair $(Z_i,W_j)$ determines the block membership of cell $(i,j)$,
and all cells in block $(k,\ell)$ share the parameters
$(\mu_{k\ell},\sigma^2_{k\ell})$. While $\mathbf Z$ and $\mathbf W$ are independent, they are in general not conditionally independent given $\mathbf Y$, because the distribution of each $Y_{ij}$ depends on both $Z_i$ and $W_j$.

\subsection{Cell likelihood and observed data likelihood}\label{subsec:cell_likelihood_observed_likelihood}

Each cell is observed either exactly or through a left- or right-censored report. For each cell $(i,j)$, let $R_{ij}$ denote the reporting rule and let $D_{ij}=R_{ij}(Y_{ij})$ denote the observed report. The matrix $\mathbf D=(D_{ij})_{1\leq i\leq n,\,1\leq j\leq m}$ collects all observed reports. Let $\delta_{ij}=\mathbb I\{\text{cell $(i,j)$ is reported exactly}\}$.

If $\delta_{ij}=1$, the exact value $D_{ij}=Y_{ij}$ is observed. If $\delta_{ij}=0$, the report states that $a_{ij}<Y_{ij}<b_{ij}$, where exactly one bound is finite, and the exact value of $Y_{ij}$ is not observed. Left censoring has $a_{ij}=-\infty$ and finite $b_{ij}$, while right censoring has finite $a_{ij}$ and $b_{ij}=\infty$. Since the Gaussian distribution is continuous, the choice of open or closed endpoints does not affect the likelihood.

Write $\mathbf d=(d_{ij})_{1\leq i\leq n,\,1\leq j\leq m}$
for the observed realization of $\mathbf D$. Each $d_{ij}$ is either an
exact numerical value or a censored interval report.

The following assumption formalizes a known Type I censoring design with a nonempty range of exact reports.

\begin{Assum}[Type I censoring design]
\label{assum:type_i_censoring}
For each cell $(i,j)$, the reporting rule $R_{ij}$ is measurable. The rule, including all reporting limits defining it, is fixed before $\mathbf Y$ is observed, is known to the analyst, and does not depend on $(\mathbf Z,\mathbf W,\boldsymbol\theta)$. There exists a nonempty open interval $I_{ij}\subset\mathbb R$ such that
\[
R_{ij}(y)=y
\quad\text{for all } y\in I_{ij},
\qquad \text{ and }  \qquad
R_{ij}(y)\notin I_{ij}
\quad\text{for all } y\notin I_{ij}.
\]
\end{Assum}

Unlike a missing entry, a censored report retains an interval constraint
on the underlying measurement. Under
Assumption~\ref{assum:type_i_censoring}, the censoring design is known
and introduces no additional model parameters.

Let $\phi$ and $\Phi$ denote the standard normal density and distribution function. For a censored report ($\delta_{ij}=0$) with bounds $a_{ij}<b_{ij}$, its likelihood is given by the conditional block probability
\begin{align*}
&P_{ij,k\ell}(d_{ij};\boldsymbol\theta):=
\Pr_{\boldsymbol\theta}
\left\{a_{ij}<Y_{ij}<b_{ij}\mid Z_i=k,W_j=\ell\right\}
\\
&\quad=
\int_{a_{ij}}^{b_{ij}}
\frac{1}{\sigma_{k\ell}}
\phi\!\left(\frac{y-\mu_{k\ell}}{\sigma_{k\ell}}\right)\,dy
=
\Phi\!\left(\frac{b_{ij}-\mu_{k\ell}}{\sigma_{k\ell}}\right)
-
\Phi\!\left(\frac{a_{ij}-\mu_{k\ell}}{\sigma_{k\ell}}\right).
\end{align*}
For an exact report with observed value $d_{ij}$ (i.e., $\delta_{ij} = 1$),
the likelihood is given by the block density
$f_{ij,k\ell}(d_{ij};\boldsymbol\theta)
=\sigma_{k\ell}^{-1}\phi\{(d_{ij}-\mu_{k\ell})/\sigma_{k\ell}\}$.
Hence, the cell likelihood for block $(k,\ell)$ is
\[
\lambda_{ij,k\ell}(d_{ij};\boldsymbol\theta)
=
\begin{cases}
f_{ij,k\ell}(d_{ij};\boldsymbol\theta), & \delta_{ij}=1,\\
P_{ij,k\ell}(d_{ij};\boldsymbol\theta), & \delta_{ij}=0.
\end{cases}
\]
For left censoring, $P_{ij,k\ell}(d_{ij};\boldsymbol\theta)$ reduces to $\Phi\{(b_{ij}-\mu_{k\ell})/\sigma_{k\ell}\}$. For right censoring, it reduces to $1-\Phi\{(a_{ij}-\mu_{k\ell})/\sigma_{k\ell}\}$.

Given fixed labels $(\mathbf z,\mathbf w)$, the reported cells are conditionally independent. The conditional observed likelihood is
\begin{equation}\label{eq:fixed_label_observed_likelihood}
p_{\boldsymbol\theta}(\mathbf d\mid\mathbf z,\mathbf w)
=
\prod_{i=1}^n\prod_{j=1}^m
\lambda_{ij,z_iw_j}(d_{ij};\boldsymbol\theta).
\end{equation}
Since the row and column labels are unobserved, the observed data likelihood is
\begin{equation}\label{eq:observed_data_likelihood_revised}
\begin{aligned}
L_{\mathrm{obs}}(\boldsymbol\theta;\mathbf d)
&=
\sum_{\mathbf z\in\{1,\ldots,K\}^n}
\sum_{\mathbf w\in\{1,\ldots,L\}^m}
p_{\boldsymbol\theta}(\mathbf z,\mathbf w)
p_{\boldsymbol\theta}(\mathbf d\mid\mathbf z,\mathbf w).
\end{aligned}
\end{equation}

\subsection{Identifiability}\label{subsec:candidate_model_classes}
The observed data likelihood specifies the distribution of the reports under a given parameter value, but it does not show whether distinct parameter values can produce the same distribution. We therefore study identifiability of CGLBM.
Identifiability of the CGLBM is established under two additional
assumptions: positive class proportions and pairwise distinct Gaussian block
distributions.

\begin{Assum}[Positive class proportions]
\label{assum:positive_weights_revised}
All row and column class probabilities are positive, with $\pi_k>0$ for $k=1,\ldots,K$ and $\rho_\ell>0$ for $\ell=1,\ldots,L$.
\end{Assum}
\begin{Assum}[Distinct block distributions]
\label{assum:block_profile_separation_revised}
For any $(k,\ell)\neq(k',\ell')$, $
(\mu_{k\ell},\sigma_{k\ell}^{2})
\neq
(\mu_{k'\ell'},\sigma_{k'\ell'}^{2}).$
\end{Assum}

Assumption~\ref{assum:positive_weights_revised}
rules out zero-probability row or column classes. Such classes are absent from the population and can be removed from the model without changing the data distribution. Assumption~\ref{assum:block_profile_separation_revised}
rules out duplicate Gaussian blocks, so each block represents a distinct
distributional pattern.

{
Let $\mathbb P_{\boldsymbol\theta}^{\mathbf Y}$ and $\mathbb P_{\boldsymbol\theta}^{\mathbf D}$ denote the exact data and report distributions under $\boldsymbol\theta$. For two parameter values $\boldsymbol\theta$ and $\widetilde{\boldsymbol\theta}$ with the same class numbers $(K,L)$, write $\boldsymbol\theta\sim\widetilde{\boldsymbol\theta}$ when there are permutations $s$ of $\{1,\ldots,K\}$ and $t$ of $\{1,\ldots,L\}$ such that the following equalities hold for every $k$ and $\ell$:
\[
\widetilde\pi_{s(k)}=\pi_k,
\qquad
\widetilde\rho_{t(\ell)}=\rho_\ell,
\qquad \text{ and } \qquad
\left(
\widetilde\mu_{s(k),t(\ell)},
\widetilde\sigma^2_{s(k),t(\ell)}
\right)
=
\left(
\mu_{k\ell},
\sigma^2_{k\ell}
\right),
\]
meaning that equivalent parameter values differ only by separate permutations of the row and column labels and induce the same exact data and report distributions.

\begin{proposition}[Reporting equivalence]
\label{prop:reporting_equivalence_preservation}
Fix the class numbers $(K,L)$ and a censoring design satisfying Assumption~\ref{assum:type_i_censoring}. For any $\boldsymbol\theta$ and $\widetilde{\boldsymbol\theta}$ with positive block variances,
\[
\mathbb P_{\boldsymbol\theta}^{\mathbf D}
=
\mathbb P_{\widetilde{\boldsymbol\theta}}^{\mathbf D}
\quad\text{if and only if}\quad
\mathbb P_{\boldsymbol\theta}^{\mathbf Y}
=
\mathbb P_{\widetilde{\boldsymbol\theta}}^{\mathbf Y}.
\]
\end{proposition}
The report law and the exact data law therefore have the same observational equivalence classes. A known Type I censoring design introduces no additional nonidentifiability into the Gaussian latent block model. The proof is given in Section~\ref{app:proof_reporting_equivalence_preservation} of the Supplementary Appendix.

\begin{Cor}[Identifiability of the CGLBM]
\label{cor:cglbm_identifiability}
Fix the class numbers $(K,L)$ and suppose $n\geq2$ and $m\geq2$. Let $\boldsymbol\theta$ and $\widetilde{\boldsymbol\theta}$ satisfy Assumptions~\ref{assum:positive_weights_revised} and~\ref{assum:block_profile_separation_revised} and have positive block variances. Under Assumption~\ref{assum:type_i_censoring},
\[
\mathbb P_{\boldsymbol\theta}^{\mathbf D}
=
\mathbb P_{\widetilde{\boldsymbol\theta}}^{\mathbf D}
\quad\text{if and only if}\quad
\boldsymbol\theta\sim\widetilde{\boldsymbol\theta}.
\]
\end{Cor}

The result first uses finite-mixture identifiability
for distinct univariate Gaussian distributions \citep{teicher1963} to
match the block distributions across parameterizations. The joint
distributions of cell pairs sharing a row or a column then reveal which
matched blocks have the same row class and which have the same column
class. This forces the block matching to factor into separate
permutations of the row and column labels. The proof is given in
Section~\ref{app:proof_cglbm_identifiability} of the Supplementary Appendix.}

\section{Estimation and model selection}\label{sec:estimation_fixed_candidate_model}
For each candidate model $M=(K,L)$, we predict the row and column memberships and estimate the model parameters using a common free energy objective.
Model selection then uses a BIC approximation to the integrated
classification likelihood to determine the row and column class numbers.

\subsection{A common objective for hard and soft memberships}\label{subsec:classification_variational_criteria}

Let $\boldsymbol\tau=(\tau_{ik})$ be the $n\times K$ row membership matrix and let $\boldsymbol\omega=(\omega_{j\ell})$ be the $m\times L$ column membership matrix. Define $\boldsymbol\tau_i=(\tau_{i1},\ldots,\tau_{iK})^\intercal$ and $\boldsymbol\omega_j=(\omega_{j1},\ldots,\omega_{jL})^\intercal$. 

For a positive integer $q$, let
$\Delta_q=\{\mathbf u=(u_1,\ldots,u_q)^\intercal\in[0,1]^q:
\sum_{a=1}^q u_a=1\}$ and
$\mathcal V_q=\{\mathbf e_1,\ldots,\mathbf e_q\}$,
where $\mathbf e_a$ denotes the $a$th standard basis vector in
$\mathbb R^q$.
We consider four membership configurations
$\mathfrak c\in\{\mathsf{HH},\mathsf{HS},\mathsf{SH},\mathsf{SS}\}$,
where the first and second letters specify the row and column membership
domains, respectively, and $\mathsf H$ and $\mathsf S$ denote hard and
soft memberships. The admissible membership sets are
\[
\begin{aligned}
\mathcal A_{\mathsf{HH}}&=\mathcal V_K^n\times\mathcal V_L^m,
&
\mathcal A_{\mathsf{HS}}&=\mathcal V_K^n\times\Delta_L^m,\\
\mathcal A_{\mathsf{SH}}&=\Delta_K^n\times\mathcal V_L^m,
&
\mathcal A_{\mathsf{SS}}&=\Delta_K^n\times\Delta_L^m.
\end{aligned}
\]

The observed data likelihood in Equation~\eqref{eq:observed_data_likelihood_revised} sums over $K^nL^m$ joint label configurations. Since every cell couples one row label and one column label, the exact posterior $p_{\boldsymbol\theta}(\mathbf Z,\mathbf W\mid\mathbf d)$ does not generally factorize and direct maximization is infeasible at the matrix sizes of interest \citep{govaert2005,biernacki2023}.

For any feasible pair
$(\boldsymbol\tau,\boldsymbol\omega)$, let
$q_{\boldsymbol\tau,\boldsymbol\omega}(\mathbf z,\mathbf w)
=\prod_{i=1}^n\tau_{i,z_i}\prod_{j=1}^m\omega_{j,w_j}$
denote the factorized probability mass function on the latent label
vectors $(\mathbf Z,\mathbf W)$.
Under this distribution, $\Pr_{q_{\boldsymbol\tau,\boldsymbol\omega}}(Z_i=k)=\tau_{ik}$ and $\Pr_{q_{\boldsymbol\tau,\boldsymbol\omega}}(W_j=\ell)=\omega_{j\ell}$. Hence, $\tau_{ik}\omega_{j\ell}$ is the membership weight assigned to cell $(i,j)$ in block $(k,\ell)$.

Define the free energy objective
\begin{equation}\label{eq:free_energy_definition}
\mathcal J(\boldsymbol\tau,\boldsymbol\omega,\boldsymbol\theta;\mathbf d)
=
\E_{q_{\boldsymbol\tau,\boldsymbol\omega}}
\left[\log p_{\boldsymbol\theta}(\mathbf d,\mathbf Z,\mathbf W)\right]
-
\E_{q_{\boldsymbol\tau,\boldsymbol\omega}}
\left[\log q_{\boldsymbol\tau,\boldsymbol\omega}(\mathbf Z,\mathbf W)\right].
\end{equation}
Following the free energy formulation of the EM algorithm \citep{dempster1977,neal1998}, $\log L_{\mathrm{obs}}(\boldsymbol\theta;\mathbf d)$ is the sum of $\mathcal J(\boldsymbol\tau,\boldsymbol\omega,\boldsymbol\theta;\mathbf d)$ and the Kullback-Leibler divergence from $q_{\boldsymbol\tau,\boldsymbol\omega}$ to the posterior $p_{\boldsymbol\theta}(\mathbf Z,\mathbf W\mid\mathbf d)$. Hence, $\mathcal J$ is a lower bound on the observed log likelihood, with equality when $q_{\boldsymbol\tau,\boldsymbol\omega}$ equals the posterior. Let $\Theta_M$ be the parameter space with positive class proportions, block means in $\mathbb R$, and $\sigma_{k\ell}^2 \geq \sigma_{\min}^2 > 0$ for every block, where
$\sigma_{\min}^2$ is a fixed variance floor that keeps block variances away from zero. For configuration $\mathfrak c$, the optimization problem is therefore
\[
\sup_{\substack{
(\boldsymbol\tau,\boldsymbol\omega)\in\mathcal A_{\mathfrak c}\\
\boldsymbol\theta\in\Theta_M}}
\mathcal J(\boldsymbol\tau,\boldsymbol\omega,\boldsymbol\theta;\mathbf d),
\qquad
\mathfrak c\in\{\mathsf{HH},\mathsf{HS},\mathsf{SH},\mathsf{SS}\}.
\]

Using the membership probabilities $\tau_{ik}$ and $\omega_{j\ell}$, the cell block weights $\tau_{ik}\omega_{j\ell}$, and the model factorization, Equation~\eqref{eq:free_energy_definition} can be written as
\begin{align}
\label{eq:unified_membership_objective}
\mathcal J(\boldsymbol\tau,\boldsymbol\omega,\boldsymbol\theta;\mathbf d)
={}&
\sum_{i=1}^n\sum_{k=1}^K \tau_{ik}\log\pi_k
+
\sum_{j=1}^m\sum_{\ell=1}^L \omega_{j\ell}\log\rho_\ell
\nonumber\\
&+
\sum_{i=1}^n\sum_{j=1}^m\sum_{k=1}^K\sum_{\ell=1}^L
\tau_{ik}\omega_{j\ell}
\log\lambda_{ij,k\ell}(d_{ij};\boldsymbol\theta)
\nonumber\\
&-
\sum_{i=1}^n\sum_{k=1}^K\tau_{ik}\log\tau_{ik}
-
\sum_{j=1}^m\sum_{\ell=1}^L\omega_{j\ell}\log\omega_{j\ell}.
\end{align}
The final two sums are the row and column entropies. Under the convention $0\log 0=0$, every simplex vertex has zero entropy. \(\mathsf{HH}\) therefore has no entropy contribution, \(\mathsf{HS}\) retains only the column entropy, \(\mathsf{SH}\) retains only the row entropy, and \(\mathsf{SS}\) retains both. For fixed $\boldsymbol\theta$ and hard labels $(\mathbf z,\mathbf w)$, the $\mathsf{HH}$ objective is therefore the classification log likelihood $\log p_{\boldsymbol\theta}(\mathbf d,\mathbf z,\mathbf w)$. Since the column labels are conditionally independent given the data and the row labels, maximizing the $\mathsf{HS}$ objective over $\boldsymbol\omega$ gives the partial marginal log likelihood $\log p_{\boldsymbol\theta}(\mathbf d,\mathbf z)$. Similarly, maximizing the $\mathsf{SH}$ objective over $\boldsymbol\tau$ gives $\log p_{\boldsymbol\theta}(\mathbf d,\mathbf w)$. Evaluating Equation~\eqref{eq:unified_membership_objective} only requires $nmKL$ cell block terms.

The \(\mathsf{HH}\) membership domain is contained in both the \(\mathsf{HS}\) and \(\mathsf{SH}\) domains,
and each is contained in the \(\mathsf{SS}\) domain. The \(\mathsf{HS}\) and \(\mathsf{SH}\) domains are not
nested. Hence, the supremum of $\mathcal J$ is nondecreasing along these
inclusions and cannot serve as a fair criterion for selecting among the
four configurations.

\subsection{Gaussian block updates and boundary conditions}\label{subsec:weighted_censored_gaussian_block_update}

Since the block means and variances enter Equation~\eqref{eq:unified_membership_objective} only through its cell likelihood term, the four membership configurations use the same block update and differ only in the cell block weights $\gamma_{ij,k\ell}=\tau_{ik}\omega_{j\ell}$.
For fixed weights, we update the block means and variances to increase the weighted observed log likelihood
\[
\ell_\gamma(\boldsymbol\theta)
=
\sum_{i=1}^n\sum_{j=1}^m\sum_{k=1}^K\sum_{\ell=1}^L
\gamma_{ij,k\ell}\log\lambda_{ij,k\ell}(d_{ij};\boldsymbol\theta),
\]
subject to $\sigma_{k\ell}^2\geq\sigma_{\min}^2$. Since each censored report contributes the normal probability of its interval, the maximizer of $\ell_\gamma$ has no closed form. We therefore take one EM step that treats the underlying measurements $Y_{ij}$ as missing data \citep{lee2012}.

Define the weighted EM auxiliary function
\begin{equation}\label{eq:weighted_Q_definition_new}
Q_t(\boldsymbol\theta)
=
\sum_{i=1}^n\sum_{j=1}^m\sum_{k=1}^K\sum_{\ell=1}^L
\gamma_{ij,k\ell}
\E_{\boldsymbol\theta^{(t)}}
\left[
\log p_{\boldsymbol\theta}
(Y_{ij}\mid Z_i=k,W_j=\ell)
\mathrel{\Big|}
D_{ij}=d_{ij},Z_i=k,W_j=\ell
\right].
\end{equation}
Since the Gaussian complete data log likelihood is quadratic in $Y_{ij}$, the auxiliary function reduces, up to an additive constant, to
\[
Q_t(\boldsymbol\theta)
=-\frac12
\sum_{i=1}^n\sum_{j=1}^m\sum_{k=1}^K\sum_{\ell=1}^L
\gamma_{ij,k\ell}
\left[
\log\sigma_{k\ell}^2
+\frac{
\overline{y^2}_{ij,k\ell}^{(t)}
-2\mu_{k\ell}\overline y_{ij,k\ell}^{(t)}
+\mu_{k\ell}^2
}{\sigma_{k\ell}^2}
\right],
\]
where $\overline y_{ij,k\ell}^{(t)}=\E_{\boldsymbol\theta^{(t)}}(Y_{ij}\mid D_{ij}=d_{ij},Z_i=k,W_j=\ell)$ and $\overline{y^2}_{ij,k\ell}^{(t)}=\E_{\boldsymbol\theta^{(t)}}(Y_{ij}^2\mid D_{ij}=d_{ij},Z_i=k,W_j=\ell)$. Hence, $Q_t$ depends on the reports only through these two conditional moments and separates across blocks.

Under the current block parameters,
$Y_{ij}\mid Z_i=k,W_j=\ell
\sim\normal\!\left(\mu_{k\ell}^{(t)},(\sigma_{k\ell}^2)^{(t)}\right)$.
For an exact report,
$\overline y_{ij,k\ell}^{(t)}=d_{ij}$ and
$\overline{y^2}_{ij,k\ell}^{(t)}=d_{ij}^2$.
For a censored report with bounds $a_{ij}<b_{ij}$, conditioning on
the report restricts this Gaussian distribution to $(a_{ij},b_{ij})$.
Define
$\widetilde a_{ij,k\ell}^{(t)}
=(a_{ij}-\mu_{k\ell}^{(t)})/\sigma_{k\ell}^{(t)}$
and
$\widetilde b_{ij,k\ell}^{(t)}
=(b_{ij}-\mu_{k\ell}^{(t)})/\sigma_{k\ell}^{(t)}$.
Applying the first two truncated normal moment formulas used in the censored Gaussian EM algorithm of \citet{lee2012} gives
\begin{align}
\overline y_{ij,k\ell}^{(t)}
&=
\mu_{k\ell}^{(t)}
+\sigma_{k\ell}^{(t)}
\frac{
\phi(\widetilde a_{ij,k\ell}^{(t)})-
\phi(\widetilde b_{ij,k\ell}^{(t)})
}{
P_{ij,k\ell}(d_{ij};\boldsymbol\theta^{(t)})},
\label{eq:first_truncated_moment_new}\\
\overline{y^2}_{ij,k\ell}^{(t)}
&=
(\mu_{k\ell}^{(t)})^2
+2\mu_{k\ell}^{(t)}\sigma_{k\ell}^{(t)}
\frac{
\phi(\widetilde a_{ij,k\ell}^{(t)})-
\phi(\widetilde b_{ij,k\ell}^{(t)})
}{
P_{ij,k\ell}(d_{ij};\boldsymbol\theta^{(t)})}
+
(\sigma_{k\ell}^{2})^{(t)}
\left\{
1+
\frac{
\widetilde a_{ij,k\ell}^{(t)}\phi(\widetilde a_{ij,k\ell}^{(t)})-
\widetilde b_{ij,k\ell}^{(t)}\phi(\widetilde b_{ij,k\ell}^{(t)})
}{
P_{ij,k\ell}(d_{ij};\boldsymbol\theta^{(t)})}
\right\}.
\label{eq:second_truncated_moment_new}
\end{align}
For an infinite endpoint, we set $\phi(\pm\infty)=0$ and $\pm\infty\cdot\phi(\pm\infty)=0$, which are the limits of these terms.

\begin{proposition}[Weighted Gaussian block update under censoring]\label{prop:weighted_censored_gaussian_update}
Define the effective block mass
$N_{k\ell}=\sum_{i=1}^n\sum_{j=1}^m\gamma_{ij,k\ell}$.
Assume $N_{k\ell}>0$ and $(\sigma_{k\ell}^2)^{(t)}\geq\sigma_{\min}^2>0$ for every block, and assume $\lambda_{ij,k\ell}(d_{ij};\boldsymbol\theta^{(t)})>0$ whenever $\gamma_{ij,k\ell}>0$. The constrained maximizer of $Q_t(\boldsymbol\theta)$ with respect to the Gaussian block parameters is
\begin{align}
\mu_{k\ell}^{(t+1)}
&=
\frac{
\sum_{i=1}^n\sum_{j=1}^m
\gamma_{ij,k\ell}\overline y_{ij,k\ell}^{(t)}
}{N_{k\ell}},
\label{eq:weighted_mean_update_new}\\
(\sigma_{k\ell}^2)^{(t+1)}
&=
\max\left\{
\frac{
\sum_{i=1}^n\sum_{j=1}^m
\gamma_{ij,k\ell}
\left[
\overline{y^2}_{ij,k\ell}^{(t)}
-2\mu_{k\ell}^{(t+1)}\overline y_{ij,k\ell}^{(t)}
+(\mu_{k\ell}^{(t+1)})^2
\right]
}{N_{k\ell}},
\sigma_{\min}^2
\right\}.
\label{eq:weighted_variance_update_new}
\end{align}
These updates satisfy
\begin{equation}\label{eq:block_update_monotonicity}
\ell_\gamma(\boldsymbol\theta^{(t+1)})\geq\ell_\gamma(\boldsymbol\theta^{(t)}).
\end{equation}
\end{proposition}

The proof is given in Section~\ref{app:proof_weighted_censored_gaussian_update} of the Supplementary Appendix.

While Proposition~\ref{prop:weighted_censored_gaussian_update} guarantees that the updates do not decrease $\ell_\gamma$, a finite maximizer does not exist under complete one-sided censoring of a block or, without the variance floor, under Gaussian variance collapse.

\begin{Lem}[Complete one-sided censoring]
\label{lem:complete_left_censoring}
Fix a block $(k,\ell)$ with $N_{k\ell}>0$. If every report assigned positive weight to this block is left censored at a finite threshold $b_{ij}$, then
\[
\sum_{i=1}^n\sum_{j=1}^m
\gamma_{ij,k\ell}
\log\Phi\!\left\{
\frac{b_{ij}-\mu_{k\ell}}{\sigma_{k\ell}}
\right\}
\]
has supremum zero over $\mu_{k\ell}\in\mathbb R$ and $\sigma_{k\ell}^2\geq\sigma_{\min}^2$, and no finite maximizer. The corresponding result holds when every report assigned positive weight is right censored.
\end{Lem}

The proof is given in Section~\ref{app:proof_complete_left_censoring} of the Supplementary Appendix. The unified GEM procedure therefore discards a run before the Gaussian update whenever all reports assigned positive weight to a block are censored in the same direction.

Gaussian variance collapse makes the likelihood unbounded as a block variance approaches zero \citep{hathaway1985}. The constraint $\sigma_{k\ell}^2\geq\sigma_{\min}^2$ for every block prevents this.

\subsection{Unified coordinate updates and the generalized EM (GEM) algorithm}\label{subsec:unified_membership_gem}

The unified GEM procedure alternates coordinate updates for the row and column memberships with an E-step for censored measurements and a generalized M-step for the model parameters. In a hard membership dimension, the coordinate objective is maximized over the simplex vertices and gives a classification update. In a soft membership dimension, it is maximized over the full simplex and gives a variational E-update.

Let a superscript $(t)$ indicate the value of a quantity at iteration $t$. For a fixed candidate model $M$ and membership configuration $\mathfrak c$, one run of the unified GEM procedure proceeds through the following steps.

\noindent\textbf{(1) Initialization.} Initialize $(\boldsymbol\tau^{(0)},\boldsymbol\omega^{(0)})\in\mathcal A_{\mathfrak c}$ and $\boldsymbol\theta^{(0)}\in\Theta_M$ with positive row class, column class, and block masses. Compute $\mathcal J^{(0)}=\mathcal J(\boldsymbol\tau^{(0)},\boldsymbol\omega^{(0)},\boldsymbol\theta^{(0)};\mathbf d)$.

\noindent\textbf{(2) Row membership update.} Evaluate $\log\lambda_{ij,k\ell}(d_{ij};\boldsymbol\theta^{(t)})$ for all cells and blocks. If any computed log interval probability for a
censored report is not finite, discard the current run and restart from
Step~1. Given $\boldsymbol\omega^{(t)}$ and $\boldsymbol\theta^{(t)}$, define the row scores
\[
A_{ik}^{(t)}
=
\log\pi_k^{(t)}
+
\sum_{j=1}^m\sum_{\ell=1}^L
\omega_{j\ell}^{(t)}
\log\lambda_{ij,k\ell}(d_{ij};\boldsymbol\theta^{(t)}).
\]
The row memberships are updated by
\[
\tau_{ik}^{(t+1)}
=
\begin{cases}
\ind\!\left\{
 k=\operatorname*{arg\,max}_{1\leq k'\leq K}A_{ik'}^{(t)}
\right\},
& \mathfrak c\in\{\mathsf{HH},\mathsf{HS}\},\\[8pt]
\displaystyle
\frac{\exp\{A_{ik}^{(t)}\}}
{\sum_{k'=1}^K\exp\{A_{ik'}^{(t)}\}},
& \mathfrak c\in\{\mathsf{SH},\mathsf{SS}\}.
\end{cases}
\]
A fixed deterministic rule resolves ties in the classification update.

\noindent\textbf{(3) Column membership update.} Given $\boldsymbol\tau^{(t+1)}$ and $\boldsymbol\theta^{(t)}$, define the column scores
\[
B_{j\ell}^{(t)}
=
\log\rho_\ell^{(t)}
+
\sum_{i=1}^n\sum_{k=1}^K
\tau_{ik}^{(t+1)}
\log\lambda_{ij,k\ell}(d_{ij};\boldsymbol\theta^{(t)}).
\]
The column memberships are updated by
\[
\omega_{j\ell}^{(t+1)}
=
\begin{cases}
\ind\!\left\{
 \ell=\operatorname*{arg\,max}_{1\leq \ell'\leq L}B_{j\ell'}^{(t)}
\right\},
& \mathfrak c\in\{\mathsf{HH},\mathsf{SH}\},\\[8pt]
\displaystyle
\frac{\exp\{B_{j\ell}^{(t)}\}}
{\sum_{\ell'=1}^L\exp\{B_{j\ell'}^{(t)}\}},
& \mathfrak c\in\{\mathsf{HS},\mathsf{SS}\}.
\end{cases}
\]
The same deterministic rule resolves ties in the classification update. If any updated row class, column class, or block mass is numerically zero, discard the current run and restart from Step 1.

\noindent\textbf{(4) E-step for censored measurements.} Set $\gamma_{ij,k\ell}^{(t+1)}=\tau_{ik}^{(t+1)}\omega_{j\ell}^{(t+1)}$ for every cell and block. If a block has positive mass and all of its positive weight reports are left censored or all are right censored, discard the current run and restart from Step 1. Otherwise, compute the conditional moments $\overline y_{ij,k\ell}^{(t)}$ and $\overline{y^2}_{ij,k\ell}^{(t)}$ under the current block parameters, as described in Section~\ref{subsec:weighted_censored_gaussian_block_update}.

\noindent\textbf{(5) Generalized M-step.}
Update the class proportions by
$\pi_k^{(t+1)}
=\frac{1}{n}\sum_{i=1}^n\tau_{ik}^{(t+1)}$
and
$\rho_\ell^{(t+1)}
=\frac{1}{m}\sum_{j=1}^m\omega_{j\ell}^{(t+1)}$.
Using the updated weights and the conditional
moments from the E-step,
Proposition~\ref{prop:weighted_censored_gaussian_update} gives the
updates of $\mu_{k\ell}^{(t+1)}$ and
$(\sigma_{k\ell}^2)^{(t+1)}$.

Since the row and column membership updates maximize \(\mathcal J\) with the other quantities fixed, they do not decrease \(\mathcal J\). The same holds for the class proportion update. Proposition~\ref{prop:weighted_censored_gaussian_update} shows that the Gaussian block update does not decrease \(\mathcal J\). Therefore, an iteration that is not discarded does not decrease \(\mathcal J\). Under the variance floor, the exact-report log densities are bounded above. The censored-report and class-proportion terms are nonpositive, and the entropy terms are bounded. Thus, \(\mathcal J\) is bounded above, so the sequence of objective values converges.

After Step 5, compute
$\mathcal J^{(t+1)}
=\mathcal J(\boldsymbol\tau^{(t+1)},
\boldsymbol\omega^{(t+1)},
\boldsymbol\theta^{(t+1)};
\mathbf d)$.
A decrease in $\mathcal J$ beyond numerical tolerance discards the current run. The run terminates when
\[
\left|\mathcal J^{(t+1)}-\mathcal J^{(t)}\right|
\leq
\varepsilon\left\{1+\left|\mathcal J^{(t)}\right|\right\}
\]
and the labels in every hard membership dimension are unchanged from the previous iteration. Steps 2 through 5 are repeated until this rule holds or $T_{\max}$ iterations have been completed.

At termination, set the maximum a posteriori (MAP) labels to
$\widehat z_i=\argmax_{1\leq k\leq K}\widehat\tau_{ik}$
and
$\widehat w_j=\argmax_{1\leq\ell\leq L}\widehat\omega_{j\ell}$.
For hard memberships, this projection returns the fitted labels directly. If either partition omits a declared class, discard the run. Otherwise, store it as a valid run.

Since the optimization problem is nonconvex, the procedure seeks $N_{\mathrm{start}}$ valid runs for each fixed model from at most $N_{\max}$ initializations. If at least one valid run is obtained, the run with the largest final value of $\mathcal J$ is retained. Otherwise, the procedure returns no fit for that model.

All likelihood and membership calculations use stable log-scale evaluations.
Interval probabilities use log distribution function differences.
To avoid loss of precision in right-tail calculations, we use log survival
functions. In softmax calculations, we subtract the largest score before
exponentiation to prevent overflow.

\subsection{Selection of row and column class numbers}\label{subsec:selection_row_column_class_numbers}

Since the observed data likelihood cannot be evaluated at the matrix sizes of interest (Section~\ref{subsec:classification_variational_criteria}), we select the class numbers with the integrated classification likelihood (ICL), which evaluates the reported data together with the fitted row and column labels. The ICL was introduced for mixture model clustering
\citep{biernacki2000} and extended to latent block models for categorical and
Gaussian data \citep{keribin2015,lomet2018}. We use a
BIC approximation to this criterion.

Let $\mathcal M_{\mathrm{cand}}\subset\mathbb N^2$ denote the finite set of candidate class-number pairs $M=(K,L)$. The same membership configuration and fitting settings are used for every candidate. For each $M$, write $\widehat{\boldsymbol\theta}_M$ for the parameter estimate from the run retained in Section~\ref{subsec:unified_membership_gem} and $\widehat{\mathbf z}_M,\widehat{\mathbf w}_M$ for the corresponding MAP labels. Let $\nu_M$ be a proper prior density on $\Theta_M$. Holding the MAP labels fixed, the integrated classification likelihood is
\begin{equation}\label{eq:integrated_classification_likelihood}
\operatorname{ICL}(M)
=
\log
\int_{\Theta_M}
p_{\boldsymbol\theta}
\left(
\mathbf d,
\widehat{\mathbf z}_M,
\widehat{\mathbf w}_M
\right)
\nu_M(\boldsymbol\theta)
\,d\boldsymbol\theta.
\end{equation}

Under the regularity conditions stated in Section~\ref{app:icl_bic_derivation} of the Supplementary Appendix, a Laplace approximation of Equation~\eqref{eq:integrated_classification_likelihood} equals the maximized classification log likelihood minus a logarithmic penalty for each parameter group, up to bounded terms. For computation, the maximized classification log likelihood in this approximation is replaced by the plug-in value $\widehat\ell_{\mathrm{cl}}(M)=\log p_{\widehat{\boldsymbol\theta}_M}(\mathbf d,\widehat{\mathbf z}_M,\widehat{\mathbf w}_M)$ at the fitted GEM estimate. By Equations~\eqref{eq:label_prior} and~\eqref{eq:fixed_label_observed_likelihood}, this quantity combines the fitted row- and column-class probabilities with the cell report likelihoods. For soft memberships, the parameters are not refitted at the MAP labels. Omitting the bounded terms gives the implemented BIC-based score
\begin{equation}\label{eq:censored_icl_bic}
\operatorname{ICL}_{\mathrm{BIC}}(M)
=
\widehat\ell_{\mathrm{cl}}(M)
-
\frac{K-1}{2}\log n
-
\frac{L-1}{2}\log m
-
\frac{2KL}{2}\log(nm).
\end{equation}

The three penalties correspond to the $K-1$ row-proportion parameters, the $L-1$ column-proportion parameters, and the $2KL$ Gaussian block parameters, whose information grows at rates $n$, $m$, and $nm$, respectively. Under Assumption~\ref{assum:type_i_censoring}, the known reporting limits introduce no additional parameters. Censoring changes the block information but preserves its $nm$ order under the nondegeneracy condition in Section~\ref{app:icl_bic_derivation}. The logarithmic penalty form therefore coincides with that in the Gaussian latent block criterion of \citet{lomet2018}. Among candidates with valid fits, we select $\widehat M=(\widehat K,\widehat L)\in\argmax_{M\in\mathcal M_{\mathrm{cand}}}\operatorname{ICL}_{\mathrm{BIC}}(M)$.

\subsection{Fisher information and approximate standard errors}
\label{subsec:fisher_information_under_censoring}

For the selected model, we compute approximate standard errors for the fitted block means from the Fisher information that each report carries about its block mean and variance, treating the selected class numbers and fitted memberships as fixed.

\begin{Lem}[Fisher information under lower and upper reporting limits]
\label{lem:left_censor_fisher_information}
Let $Y\sim\normal(\mu,\sigma^2)$ with $\sigma^2>0$, and let
$-\infty\leq c_L<c_U\leq\infty$ be fixed reporting limits.
The report records $(-\infty,c_L]$ when $Y\leq c_L$ and
$[c_U,\infty)$ when $Y\geq c_U$.
For $c_L<Y<c_U$, it records the exact value $Y$.
Write
$\alpha=(c_L-\mu)/\sigma$,
$\beta=(c_U-\mu)/\sigma$,
$p_L=\Phi(\alpha)$,
$p_R=1-\Phi(\beta)$, and
$q=\Phi(\beta)-\Phi(\alpha)$.
One report has Fisher information for $(\mu,\sigma^2)$
\begin{equation}\label{eq:lower_upper_fisher_information}
\mathcal I(\mu,\sigma^2;c_L,c_U)
=
\begin{pmatrix}
A(\alpha,\beta)/\sigma^2
&
B(\alpha,\beta)/(2\sigma^3)
\\[4pt]
B(\alpha,\beta)/(2\sigma^3)
&
C(\alpha,\beta)/(4\sigma^4)
\end{pmatrix},
\end{equation}
where
\[
\begin{aligned}
A(\alpha,\beta)
&=
q+\alpha\phi(\alpha)-\beta\phi(\beta)
+\frac{\phi(\alpha)^2}{p_L}
+\frac{\phi(\beta)^2}{p_R},\\[4pt]
B(\alpha,\beta)
&=
(\alpha^2+1)\phi(\alpha)-(\beta^2+1)\phi(\beta)
+
\frac{\alpha\phi(\alpha)^2}{p_L}
+\frac{\beta\phi(\beta)^2}{p_R},\\[4pt]
C(\alpha,\beta)
&=
2q+(\alpha^3+\alpha)\phi(\alpha)
-(\beta^3+\beta)\phi(\beta)
+
\frac{\alpha^2\phi(\alpha)^2}{p_L}
+\frac{\beta^2\phi(\beta)^2}{p_R}.
\end{aligned}
\]
When a reporting limit is infinite, every term containing $\phi$ at that limit is set to its limiting value, which is zero.
The information matrix is positive definite.
\end{Lem}

For left censoring at a finite threshold $c$,
take $c_L=c$ and $c_U=\infty$.
The information entries in this special case follow from the
Type I Tobit score given by \citet{amemiya1984}.
For fixed $(\mu,\sigma^2)$, the information is nonincreasing in the
Loewner order as the censoring probability $p=\Phi(\alpha)$ increases.
It converges to the exact Gaussian information as $p\to0$ and to the
zero matrix as $p\to1$.
The proof is given in
Section~\ref{app:proof_left_censor_fisher_information} of the Supplementary Appendix.

By contrast, if only the censoring indicator $\mathbb{I}\{Y\leq c\}$ were observed, its Fisher information would have rank one, since its distribution depends on $(\mu,\sigma^2)$ only through $(c-\mu)/\sigma$ (Section~\ref{app:censoring_indicator_information} of the Supplementary Appendix).

For each cell $(i,j)$, let $
\mathcal I_{ij,k\ell}
\left(
\widehat{\mu}_{k\ell},
\widehat{\sigma}_{k\ell}^{2}
\right)$
denote the information from the preceding lemma under the known reporting limits for that cell, evaluated at the fitted parameters of block $(k,\ell)$. With $
\widehat{\gamma}_{ij,k\ell}
=
\widehat{\tau}_{ik}\widehat{\omega}_{j\ell}$,
the information matrix for block $(k,\ell)$ is
\[
\widehat{\mathcal I}_{k\ell}
=
\sum_{i=1}^{n}
\sum_{j=1}^{m}
\widehat{\gamma}_{ij,k\ell}
\mathcal I_{ij,k\ell}
\left(
\widehat{\mu}_{k\ell},
\widehat{\sigma}_{k\ell}^{2}
\right).
\]
Approximating the covariance matrix of the fitted block parameters by
$\widehat{\mathcal I}_{k\ell}^{-1}$, we use
$\widehat{\operatorname{se}}(\widehat\mu_{k\ell})
=\{(\widehat{\mathcal I}_{k\ell}^{-1})_{11}\}^{1/2}$.

\section{Simulation studies}\label{sec:simulation_studies}

We compared CGLBM with procedures that complete censored reports before fitting a Gaussian latent block model in Experiment~I, which examined partition recovery and parameter estimation under the four membership configurations. Experiment~II examined the ICL criterion for selecting the numbers of row and column classes.

Both experiments generated matrices with \(n=120\) rows and \(m=60\) columns from Gaussian latent block models. In each replicate, the true label vectors \(\mathbf z^\star\) and \(\mathbf w^\star\) were generated independently by random assignment subject to fixed row class sizes \((n_1,\ldots,n_K)\) and column class sizes \((m_1,\ldots,m_L)\), with all classes nonempty. Conditional on the true labels, the latent cells were independent and
\[
Y_{ij}^\star \mid z_i^\star=k,w_j^\star=\ell
\sim
\normal\!\left(
\mu_{k\ell}^\star,
\left(\sigma_{k\ell}^\star\right)^2
\right),
\qquad
k=1,\ldots,K,
\quad
\ell=1,\ldots,L.
\]

Since Assumption~\ref{assum:type_i_censoring} requires reporting limits that are fixed before the data are observed, the lower and upper detection limits were set from independent calibration samples. Let \(\widehat F_{\mathrm{cal}}\) be the empirical distribution of the pooled calibration cells. For an overall target censoring rate \(\zeta>0\) and lower tail allocation \(\eta\), the limits were
\begin{equation}\label{eq:simulation_detection_limits}
 c_L
 =
 \widehat F_{\mathrm{cal}}^{-1}(\zeta\eta),
 \qquad
 c_U
 =
 \widehat F_{\mathrm{cal}}^{-1}\{1-\zeta(1-\eta)\}.
\end{equation}
Measurements at or below \(c_L\) were left censored, those at or above \(c_U\) were right censored, and those between the limits were reported exactly. The target \(\zeta\) was defined in the pooled calibration distribution, so realized censoring rates could vary across matrices. Section~\ref{app:simulation_common_details} of the Supplementary Appendix gives the calibration details.

\subsection{Censoring likelihood and membership configuration}\label{subsec:experiment_i_likelihood_preprocessing}

Experiment I used the true class numbers \(K=L=3\), with row class sizes of \(40,40,40\) and column class sizes of \(20,20,20\) in the balanced design.

The balanced factorial design combined the following factors:
\begin{itemize}
\item Signal strength \((a_R^\star,a_C^\star)\in
\{(0.75,0.400),(0.60,0.320),(0.35,0.185)\}\).
\item Overall target censoring rate \(\zeta\in\{0.10,0.20,0.30\}\).
\item Censoring direction \(\eta\in\{0.20,0.50,0.80\}\).
\end{itemize}
The three signal settings corresponded to strong, moderate, and weak signal, respectively. The values \(\eta=0.20,0.50,0.80\) specified right heavy, symmetric, and left heavy censoring, respectively.

Let \(\mathbf s=(-1,0,1)^\top\) and
\[
\boldsymbol\Gamma
=
\begin{pmatrix}
1 & -1 & 0\\
-1 & 0 & 1\\
0 & 1 & -1
\end{pmatrix}.
\]
The row and column sums of \(\boldsymbol\Gamma\) are zero. The true block means were $
\mu_{k\ell}^\star
=
a_R^\star s_k
+
a_C^\star s_\ell
+
0.25\sqrt{a_R^\star a_C^\star}\,\Gamma_{k\ell}.$
The first two terms control row and column separation. The interaction term allows the block means to depart from an additive row and column structure without changing the marginal main effects. The true block variances were
\[
\left\{\left(\sigma_{k\ell}^\star\right)^2\right\}_{k,\ell=1}^{3}
=
\begin{pmatrix}
0.880 & 0.978 & 1.076\\
1.027 & 0.929 & 1.057\\
0.998 & 1.096 & 0.959
\end{pmatrix}.
\]
The variance structure averaged to one, so all three signal settings used the same overall variance scale.

The balanced factorial design included 27 settings (3 signal strengths $\times$ 3 censoring rates $\times$ 3 censoring directions). For each signal strength, the design used 100 independently generated latent matrices, each reported under all nine censoring rate and direction combinations.

For the class size extension, the reference setting had balanced row and column class sizes, moderate signal strength, and 20\% symmetric censoring. Row imbalance used row class sizes of \(60,40,20\), with equal column class sizes. Column imbalance used column class sizes of \(30,20,10\), with equal row class sizes. Imbalance in both dimensions combined these two class size structures. Each setting used 100 replicates.

CGLBM was compared with the following completion procedures:
\begin{itemize}
\item \textbf{Boundary substitution.} Left and right censored reports were replaced by the corresponding detection limits \citep{lubin2004,helsel2006}.
\item \textbf{Half or double substitution.} Left-censored reports were replaced by half the lower detection limit on the original scale \citep{lubin2004,helsel2006}. Right-censored reports were replaced by twice the upper detection limit.
\item \textbf{Truncated normal imputation.} A Gaussian distribution was estimated from the exact reports, and one value was drawn for each censored report from that distribution restricted to its reported interval.
\item \textbf{Multiple imputation.} Ten completed matrices were generated using a columnwise Bayesian censored Gaussian sampler. This procedure follows the general strategy of multiple imputation followed by clustering \citep{faucheux2021}.
\end{itemize}
Completed matrices were treated as exact data when fitting the Gaussian latent block model. All five procedures were fitted under \(\mathsf{HH}\), \(\mathsf{HS}\), \(\mathsf{SH}\), and \(\mathsf{SS}\) in the balanced factorial and class size settings. Within each configuration, the five procedures used the same report matrices and ordered pool of initial row and column partitions. The configuration comparison used the same report matrices, with a separately generated ordered pool of initial partitions for each configuration. Each fit requested 20 valid starts. Numerical settings and completion details are given in Sections~\ref{app:simulation_common_details} and~\ref{app:experiment_one_design} of the Supplementary Appendix.
In a separate runtime benchmark at 20\% symmetric censoring with
balanced classes, the median time per CGLBM fit under \(\mathsf{SS}\)
was 0.819, 1.143, and 3.856 seconds under strong,
moderate, and weak signal, respectively, on a single thread
(Table~\ref{tabexperimentiruntime} in the Supplementary Appendix).

Joint partition recovery was assessed by the co-clustering adjusted Rand index (CARI), which applies the adjusted Rand index \citep{hubert1985} to the cellwise row and column label pairs \citep{robert2021}.
\[
\operatorname{CARI}
=
\operatorname{ARI}
\left[
\operatorname{vec}\!\left\{(z_i^\star,w_j^\star)\right\}_{i,j},
\operatorname{vec}\!\left\{(\widehat z_i,\widehat w_j)\right\}_{i,j}
\right].
\]
Here $\operatorname{vec}$ stacks the cellwise label pairs in a fixed common order. Row ARI and Column ARI assessed recovery in each dimension. Block mean estimation was assessed after label alignment by the block mean contrast ratio and the aligned block mean RMSE. The contrast ratio divides the estimated contrast between the blocks with the largest and smallest true means by the corresponding true contrast, so that a value below one indicates attenuation. Their definitions are given in Section~\ref{app:simulation_common_details} of the Supplementary Appendix.

Means and paired differences were first computed within each independent Monte Carlo unit and then averaged across units, with two sided 95\% Monte Carlo confidence intervals. The latent matrix was the unit for the balanced factorial and high censoring comparisons, and the matched replicate was the unit for the class size comparison. Balanced factorial summaries were computed separately for each signal
strength, and comparisons among the five procedures used \(\mathsf{SS}\). Section~\ref{app:simulation_common_details} of the Supplementary Appendix describes the matching across settings.

\begin{figure}[H]
\centering
\includegraphics[width=0.98\textwidth]{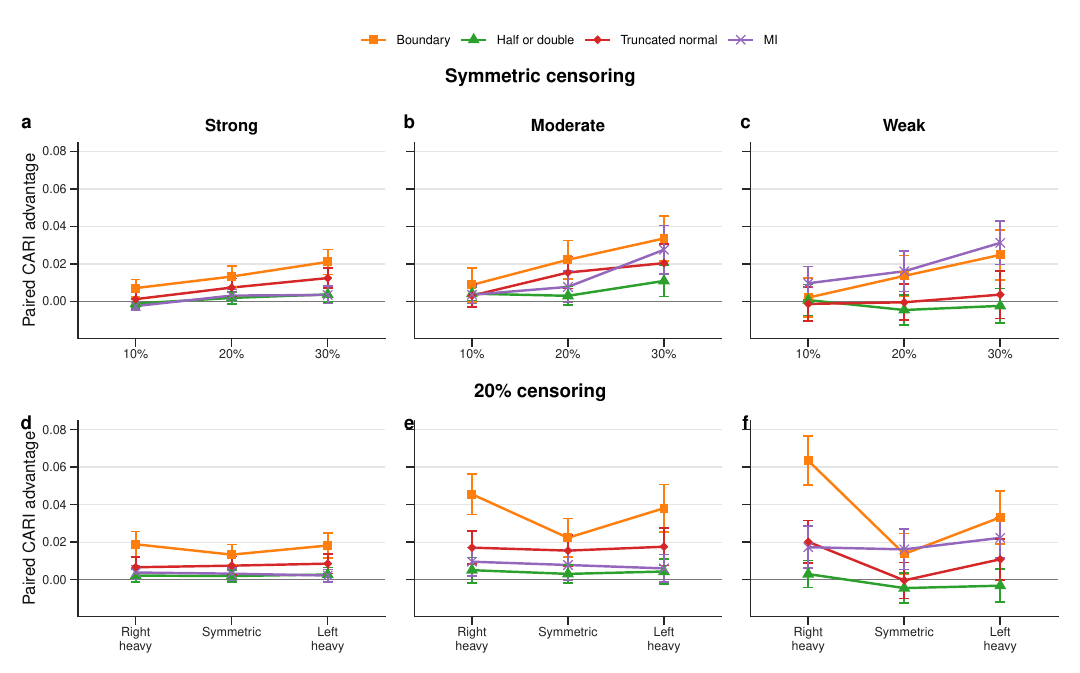}
\refstepcounter{figure}\label{figsimulationadvantage}
\par\smallskip
\begin{minipage}{0.98\textwidth}
\small\textbf{Figure \thefigure.} Paired CARI advantages for CGLBM against Boundary, Half or double, Truncated normal, and MI under \(\mathsf{SS}\). Figures~\ref{figsimulationadvantage}(a)--\ref{figsimulationadvantage}(c) show advantages by censoring rate under symmetric censoring for strong, moderate, and weak signal, respectively. Figures~\ref{figsimulationadvantage}(d)--\ref{figsimulationadvantage}(f) show advantages by censoring direction at 20\% censoring for the same signal strengths. Positive values favor CGLBM. Error bars show 95\% Monte Carlo confidence intervals.
\end{minipage}
\end{figure}

After averaging over censoring directions, the paired CARI advantages were larger at 30\% than at 10\% censoring for all four completion procedures within each signal strength (Table~\ref{tabbaselinebyfactor}, Panel B, in the Supplementary Appendix). Under moderate signal and 30\% symmetric censoring, the paired CARI advantages were 0.034 over Boundary, 0.011 over Half or double, 0.020 over Truncated normal, and 0.028 over MI. All four 95\% Monte Carlo confidence intervals were above zero (Figure~\ref{figsimulationadvantage}(b)). Across the nine censoring conditions, the advantage over Boundary was larger under moderate and weak signal than under strong signal. The advantages over Half or double and Truncated normal were largest under moderate signal, with the Half or double advantage close to zero under weak signal. The advantage over MI increased from strong to moderate to weak signal (Table~\ref{tabbaselinebyfactor}, Panel A, in the Supplementary Appendix).

At 20\% censoring, for each completion procedure and signal strength, at least one of the right-heavy and left-heavy designs produced a larger paired CARI advantage than symmetric censoring. For Boundary and Half or double, both asymmetric directions produced larger advantages than symmetric censoring at all three signal strengths (Figures~\ref{figsimulationadvantage}(d)--\ref{figsimulationadvantage}(f)). The complete $3\times3\times3$ factorial results under balanced class sizes and $\mathsf{SS}$ are shown in Figure~\ref{figfullfactorialcomparison} in the Supplementary Appendix. In the class-size extension, row imbalance gave paired CARI advantages of 0.025 over Boundary, 0.005 over Half or double, 0.016 over Truncated normal, and 0.006 over MI. The 95\% Monte Carlo confidence intervals were above zero for Boundary and Truncated normal (Table~\ref{tabbaselinebyfactor}, Panel D, in the Supplementary Appendix).

\begin{figure}[H]
\centering
\includegraphics[width=0.98\textwidth]{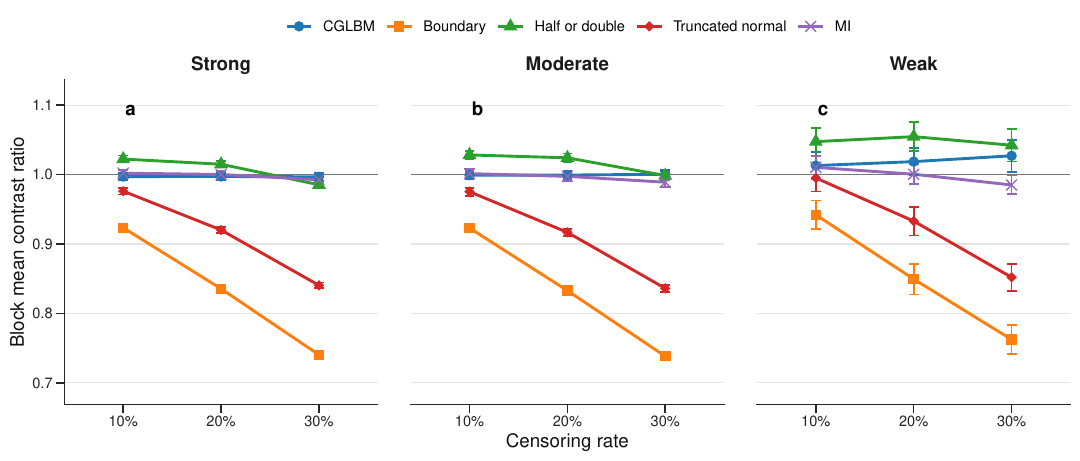}
\refstepcounter{figure}\label{figblockmeancontrast}
\par\smallskip
\begin{minipage}{0.98\textwidth}
\small\textbf{Figure \thefigure.} Block mean contrast ratios under \(\mathsf{SS}\). Figures~\ref{figblockmeancontrast}(a)--\ref{figblockmeancontrast}(c) show the estimated contrast between the blocks with the largest and smallest true means divided by the corresponding true contrast, by censoring rate under strong, moderate, and weak signal, respectively. Results are averaged over censoring directions within each latent matrix. The horizontal reference line marks one. Error bars show 95\% Monte Carlo confidence intervals.
\end{minipage}
\end{figure}

Across signal strengths and censoring rates, CGLBM gave block mean contrast ratios close to one (Figure~\ref{figblockmeancontrast}). The ratios for all four completion procedures generally decreased as the censoring rate increased, with the decreases most pronounced for Boundary and Truncated normal. MI remained close to one, although its ratios decreased as the censoring rate increased. Half or double remained mostly above one and farther from one than CGLBM. Under moderate signal, for example, the mean ratios at 10\% and 30\%
censoring were 0.92 and 0.74 for Boundary, 1.03 and 1.00 for Half or double, 0.98 and 0.84 for Truncated normal, and 1.00 and 0.99 for MI, while the corresponding CGLBM ratios were 1.00 and 1.00.

Across the nine censoring conditions within each signal strength, CGLBM had the lowest mean aligned block mean RMSE under strong and moderate signal. Under weak signal, its mean RMSE was lower than those of Boundary, Half or double, and Truncated normal, but higher than that of MI. Under moderate signal, for example, the mean RMSE was 0.042 for CGLBM, compared with 0.122 for Boundary, 0.045 for Half or double, 0.077 for Truncated normal, and 0.046 for MI
(Table~\ref{tabexponeclaimsupp}, Panel C, in the Supplementary Appendix). For CGLBM, empirical coverage of the nominal 95\% intervals for the
block means, built from the standard errors in
Section~\ref{subsec:fisher_information_under_censoring}, was approximately 0.93 under strong signal and 0.91 under moderate signal, and ranged from 0.62 to 0.65 under weak signal (Table~\ref{tabexponeclaimsupp}, Panel~D, in the Supplementary Appendix). Coverage followed joint partition recovery, which was almost exact under strong signal and poor under weak signal (mean CARI 0.965 and 0.405,
Table~\ref{tabsimulationvalues}), since the standard errors treat the fitted memberships as fixed.

Taken together, the advantage of CGLBM in joint partition recovery grew with the censoring rate against Boundary, Truncated normal, and MI at every signal strength, and CGLBM preserved the block mean contrast that Boundary and Truncated normal attenuated.

The differences among the CGLBM membership configurations increased as the signal weakened (Figure~\ref{figsimulationconfiguration} in the Supplementary Appendix). Under strong signal, the mean advantages of \(\mathsf{SS}\) over the three alternatives were close to zero. Under moderate signal, \(\mathsf{SS}\) exceeded \(\mathsf{HH}\) and \(\mathsf{SH}\) by 0.004 and 0.003, respectively, and gave similar recovery to \(\mathsf{HS}\) (Table~\ref{tabconfigurationbreakdown}, Panel A, in the Supplementary Appendix). Under weak signal, mean advantages over all three alternatives remained positive across censoring rates and censoring directions (Figures~\ref{figsimulationconfiguration}(c) and~\ref{figsimulationconfiguration}(f) in the Supplementary Appendix). Averaged over the nine censoring conditions within weak signal, the
mean paired advantages of \(\mathsf{SS}\) over \(\mathsf{HH}\), \(\mathsf{HS}\), and \(\mathsf{SH}\) were 0.049, 0.029, and 0.013, respectively. These results support \(\mathsf{SS}\) as the default membership configuration.

With \(\mathsf{SS}\) fixed, a high-censoring stress test examined how joint
partition recovery declined as the overall target censoring rate increased
from 30\% to 50\% and 75\%
(Figure~\ref{figsimulationstresstest} in the Supplementary Appendix). The design used moderate signal
strength, symmetric censoring, and balanced row and column class sizes, with
100 replicates per rate.
From 30\% to 75\% censoring, the mean CARI declined by
0.084 for CGLBM, compared with 0.124 for Boundary,
0.114 for Half or double,
0.125 for Truncated normal, and 0.345 for MI. Within each matched replicate, decline was defined as the CARI at 30\% censoring minus that at 75\% censoring. The paired differences in decline, calculated as CGLBM minus the comparison procedure, were \(-0.039\) [\(-0.061\), \(-0.017\)] for Boundary,
\(-0.030\) [\(-0.049\), \(-0.011\)] for Half or double, \(-0.040\) [\(-0.058\), \(-0.022\)] for Truncated normal, and \(-0.261\) [\(-0.278\), \(-0.244\)] for MI. All four intervals were below zero, indicating a smaller decline for
CGLBM. At 75\% censoring, the mean CARI was
0.501 for MI, compared with
0.717--0.790 for the other four procedures.

\subsection{Row and column class number selection}\label{subsec:experiment_ii_model_selection}

The main design of Experiment~II used balanced row and column class sizes and symmetric censoring, with the following factors:
\begin{itemize}
\item True class-number pair
\((K^\star,L^\star)\in\{(2,3),(3,2),(3,3)\}\).
\item Signal strength, with strong, moderate, and weak settings.
\item Overall target censoring rate
\(\zeta\in\{0,0.10,0.20,0.30\}\).
\end{itemize}

Block means contained row, column, and interaction terms, and block variances averaged to one. Signal strengths were calibrated to match the minimum row and column separations, measured by the symmetric Gaussian Kullback-Leibler divergence, across the three true class-number pairs, with the \((3,3)\) setting using the values of \((a_R^\star,a_C^\star)\) from Experiment~I (Table~\ref{tabexperimentiicalibration} in the Supplementary Appendix).

Under moderate signal and 20\% censoring, additional settings considered right heavy and left heavy censoring with balanced class sizes, or row, column, and two way imbalance with symmetric censoring. Each of the 51 settings used 100 replicates. Section~\ref{app:experiment_two_design} of the Supplementary Appendix gives the generation parameters, class sizes, and calibration details.

Since Experiment~I supported \(\mathsf{SS}\) as the default membership configuration, each dataset was fitted with \(\mathsf{SS}\) over the candidate set \(\mathcal M_{\mathrm{cand}}=\{(K,L)\mid K,L\in\{1,2,3,4,5\}\}\), using the same fitting settings as Experiment~I.

The primary outcome was the proportion of replicates in which the selected pair equaled \((K^\star,L^\star)\). Recovery of \(K^\star\) and \(L^\star\) was also recorded separately, together with the distribution of selected pairs. Recovery proportions are reported with 95\% Wilson intervals \citep{wilson1927}.

\begin{figure}[H]
\centering
\includegraphics[width=0.99\textwidth]{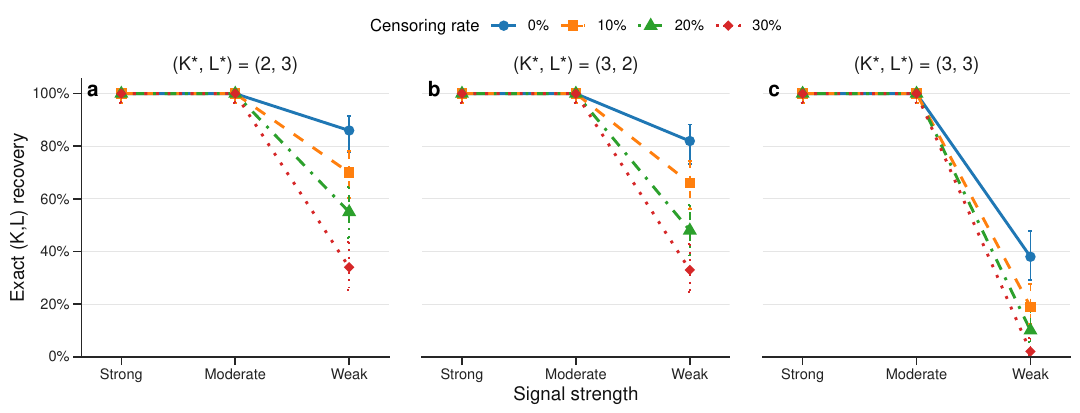}
\refstepcounter{figure}\label{figexperimentiiaccuracy}
\par\smallskip
\begin{minipage}{0.98\textwidth}
\small\textbf{Figure \thefigure.} Exact recovery of the row and column class numbers by signal strength and censoring rate. Each panel corresponds to one true class number pair. Points show exact pair recovery proportions in the balanced symmetric main design. Each point is based on 100 independently generated matrices. Error bars are 95\% Wilson intervals.
\end{minipage}
\end{figure}

In the balanced symmetric design, the ICL criterion selected the true class-number pair in every replicate under strong and moderate signal, at every censoring rate (Figure~\ref{figexperimentiiaccuracy}). Recovery was also exact under both asymmetric censoring directions at moderate signal, and in the class size extension except for one row-imbalanced \((3,3)\) replicate, which selected \((3,2)\).

Under weak signal, exact pair recovery decreased as the censoring rate increased, from 86\% without censoring to 34\% at 30\% censoring for \((2,3)\), from 82\% to 33\% for \((3,2)\), and from 38\% to 2\% for \((3,3)\) (Table~\ref{tabexperimentiifittingdiagnostics}, Panel~A, in the Supplementary Appendix). Since the design sets the column separation below the row separation at every signal strength (Table~\ref{tabexperimentiicalibration}), reduced class numbers occurred mainly in the column dimension or in both dimensions (Figure~\ref{figexperimentiifailures} in the Supplementary Appendix). Across all 5,100 datasets, no selected model contained more row or column classes than the generating model. Section~\ref{app:experiment_two_additional_results} of the Supplementary Appendix gives additional model selection results and fitting diagnostics.

\section{Discussion}\label{sec:discussion}

We proposed the censored Gaussian latent block model (CGLBM), which combines exact measurements and interval information from censored reports in a single likelihood to jointly estimate row and column memberships and block parameters. The method extends the likelihood-based treatment of censoring in finite mixture models \citep{lee2012} to coupled row and column partitions, allowing censored reports to contribute directly to block structure estimation without first completing the matrix. Unlike LBM extensions that model the missingness process \citep{frisch2022}, CGLBM uses the numerical constraints that known reporting limits impose on the underlying continuous measurements. Under a known Type I censoring design, censoring adds no nonidentifiability to the Gaussian LBM, so that the block parameters remain identifiable up to separate row and column label permutations.

In the simulations, every completion procedure fell short of CGLBM in estimating the block means or in recovering the partitions. The shortfall differed by completion rule. Boundary and Truncated normal attenuated the contrast between block means as the censoring rate increased, with block mean contrast ratios of 0.74 and 0.84, respectively, under moderate signal and 30\% censoring. Half or double overstated the contrast under weak signal. Its joint partition recovery also declined by 0.114 from 30\% to 75\% censoring, compared with 0.084 for CGLBM. While MI preserved the contrast, its joint partition recovery declined by 0.345 over the same range. CGLBM kept the contrast ratio close to one at every signal strength and censoring rate, with a value of 1.00 under moderate signal and 30\% censoring (Figure~\ref{figblockmeancontrast}). CGLBM also had the lowest block mean RMSE under strong and moderate signal. Its advantage in joint partition recovery grew with the censoring rate against Boundary, Truncated normal, and MI at every signal strength.

Since the censored reports enter the likelihood directly, the same likelihood also provides standard errors for the block means and a criterion for the class numbers. Fisher information quantifies information loss under censoring and yields approximate standard errors for fitted block means. In Experiment~I, the nominal 95\% intervals built from these standard errors had coverage of 0.93 and 0.91 under strong and moderate signal, respectively. Boundary, Half or double, and Truncated normal treat completed values as exact and thus omit completion uncertainty from standard errors. MI can combine within- and between-imputation variances using Rubin's rules, but in our columnwise procedure the latter reflects the imputation model rather than the censored block likelihood. For class number selection, we adapt the Gaussian LBM ICL approach of \citet{lomet2018} to the likelihood of censored reports. Known reporting limits introduce no additional model parameters, and under the stated regularity conditions, the BIC approximation retains the logarithmic penalty form of the Gaussian LBM. In Experiment~II, this criterion selected the true numbers of row and column classes in all but one replicate under strong and moderate signal.

The common free energy objective unifies estimation with hard and soft memberships. One GEM algorithm therefore fits all four membership configurations. We recommend $\mathsf{SS}$ as the default, since it gave the highest joint partition recovery under weak signal. The median time per $\mathsf{SS}$ fit was 0.8 to 3.9 seconds across signal strengths on a single thread for $120\times 60$ matrices.

The standard errors treat the fitted memberships as fixed, so their coverage fell below the nominal level under weak signal, where partition recovery was low. A parametric bootstrap that refits CGLBM to matrices simulated from the fitted model under the same reporting limits could carry membership uncertainty into the intervals. The interval probabilities of censored reports also rest on the Gaussian tail of each block. Drawing on censored Student-$t$ mixture models \citep{lachos2017}, a Student-$t$ extension could retain the interval likelihood and assess the sensitivity of block parameter estimates and row and column partitions to tail assumptions. CGLBM further assumes fixed, known reporting limits. When reporting limits are estimated from independent calibration data, an error model could propagate their uncertainty into inference on block parameters and partitions and assess its effect on block structure recovery. CGLBM also assumes that the reporting limits do not depend on the row or column classes. When the reporting limits depend on the latent row or column classes, for example because the classes were assayed in separate batches, the limits themselves carry information about the partition, and the reports and the limits would need to be modeled jointly.

\section*{Acknowledgements}

This work was supported by the Natural Sciences and Engineering Research Council of Canada (NSERC) through a Discovery Grant to Q.Z. (RGPIN-2023-03854). Q.Z. holds a Junior 1 research scholar award from the Fonds de recherche du Qu\'ebec. This research was enabled in part by support provided by the Digital Research Alliance of Canada (alliancecan.ca).

\section*{Data Availability Statement}

The data and reproduction materials supporting the simulation studies are openly available in the GitHub repository at \url{https://github.com/StaGill/CGLBM-reproduction}. The proposed method is implemented in the R package \texttt{CGLBM},
which is freely available at
\url{https://github.com/StaGill/CGLBM}.

\section*{Generative AI Statement}

In accordance with the Taylor \& Francis AI Policy, the authors disclose that
ChatGPT (OpenAI; versions GPT-5.6 and GPT-6) was used during manuscript
preparation to assist with language refinement and stylistic editing, as well
as to support code debugging and documentation. All scientific content,
analyses, results, interpretations, and conclusions were developed by the
authors, and all AI-assisted text and code were reviewed and verified by the
authors for accuracy and originality. The authors have reviewed the applicable
terms of use and take full responsibility for the integrity of the manuscript,
including the accuracy of its references.


\bibliographystyle{jcgs-doi}
\clearpage
\phantomsection 
\renewcommand*{\bibname}{References}

\addcontentsline{toc}{chapter}{\textbf{References}}

\bibliography{reference-arxiv}

\clearpage
\section*{Supplementary Appendix}
\appendix

\section{Supporting results for Sections 2 and 3}\label{app:supporting_results_sections2_3}
\setcounter{equation}{0}
\renewcommand{\theequation}{A\arabic{equation}}
\renewcommand{\theHequation}{A\arabic{equation}}

\subsection{Proof of Proposition~\ref{prop:reporting_equivalence_preservation}}\label{app:proof_reporting_equivalence_preservation}

\begin{proof}
Suppose $\mathbb P_{\boldsymbol\theta}^{\mathbf D}=\mathbb P_{\widetilde{\boldsymbol\theta}}^{\mathbf D}$. Let $\mathbf R(\mathbf y)=(R_{ij}(y_{ij}))_{i,j}$ and $\mathcal I=\prod_{i=1}^{n}\prod_{j=1}^{m}I_{ij}$, a nonempty open subset of $\mathbb R^{nm}$. By Assumption~\ref{assum:type_i_censoring}, $\mathbf R(\mathbf y)=\mathbf y$ for $\mathbf y\in\mathcal I$. The second condition gives $\mathbf R(\mathbf y)\in\mathcal I\Rightarrow\mathbf y\in\mathcal I$, and hence $\mathbf R(\mathbf y)=\mathbf y$. Thus, for every Borel set $A\subset\mathcal I$, $\mathbf R^{-1}(A)=A$ and
\[
\begin{aligned}
\Pr_{\boldsymbol\theta}(\mathbf Y\in A)
&=\Pr_{\boldsymbol\theta}\{\mathbf R(\mathbf Y)\in A\}\\
&=\Pr_{\widetilde{\boldsymbol\theta}}\{\mathbf R(\mathbf Y)\in A\}
=\Pr_{\widetilde{\boldsymbol\theta}}(\mathbf Y\in A).
\end{aligned}
\]

Let $p_{\boldsymbol\theta}^{Y}$ and $p_{\widetilde{\boldsymbol\theta}}^{Y}$ denote the exact data densities. Equations~\eqref{eq:label_prior} and~\eqref{eq:block_gaussian_law_revised} give
\begin{equation}\label{eq:exact_lbm_joint_density_identification}
\begin{aligned}
p_{\boldsymbol\theta}^{Y}(\mathbf y)
&=\sum_{\mathbf z\in\{1,\ldots,K\}^{n}}
\sum_{\mathbf w\in\{1,\ldots,L\}^{m}}
\left(\prod_{i=1}^{n}\pi_{z_i}\right)
\left(\prod_{j=1}^{m}\rho_{w_j}\right)\times
\prod_{i=1}^{n}\prod_{j=1}^{m}
\frac{1}{\sigma_{z_iw_j}}
\phi\!\left(\frac{y_{ij}-\mu_{z_iw_j}}{\sigma_{z_iw_j}}\right),
\end{aligned}
\end{equation}
with the corresponding expression under $\widetilde{\boldsymbol\theta}$. Equality of the probabilities on all Borel subsets of $\mathcal I$ implies $p_{\boldsymbol\theta}^{Y}=p_{\widetilde{\boldsymbol\theta}}^{Y}$ almost everywhere on $\mathcal I$. By continuity, the equality holds throughout $\mathcal I$.

Each density in Equation~\eqref{eq:exact_lbm_joint_density_identification} is a finite sum of products of Gaussian densities and is therefore real analytic in each coordinate. Fix all entries other than $y_{11}$ in their respective intervals. As functions of $y_{11}$, the two densities agree on $I_{11}$, so the univariate identity theorem extends their equality to every $y_{11}\in\mathbb R$. Apply the same argument successively to the remaining entries, allowing the previously treated entries to range over $\mathbb R$ and keeping the others in their intervals. This gives $p_{\boldsymbol\theta}^{Y}=p_{\widetilde{\boldsymbol\theta}}^{Y}$ on $\mathbb R^{nm}$, and hence $\mathbb P_{\boldsymbol\theta}^{\mathbf Y}=\mathbb P_{\widetilde{\boldsymbol\theta}}^{\mathbf Y}$.

Conversely, applying the same measurable map $\mathbf R$ to equal exact data laws gives equal report laws.
\end{proof}

\subsection{Proof of Corollary~\ref{cor:cglbm_identifiability}}\label{app:proof_cglbm_identifiability}

\begin{proof}
Suppose $\mathbb P_{\boldsymbol\theta}^{\mathbf D}=\mathbb P_{\widetilde{\boldsymbol\theta}}^{\mathbf D}$. Proposition~\ref{prop:reporting_equivalence_preservation} gives $\mathbb P_{\boldsymbol\theta}^{\mathbf Y}=\mathbb P_{\widetilde{\boldsymbol\theta}}^{\mathbf Y}$.

We first establish linear independence of distinct univariate Gaussian densities. Let $f_1,\ldots,f_q$ have pairwise distinct parameter pairs $(\nu_r,v_r)$. If $\sum_{r=1}^{q}a_rf_r(x)=0$ almost everywhere, integrating against $e^{tx}$ gives
\begin{equation}\label{eq:gaussian_density_linear_independence_mgf}
\sum_{r=1}^{q}a_r\exp\!\left(\nu_rt+\frac12v_rt^2\right)=0,
\qquad t\in\mathbb R,
\end{equation}
since every Gaussian moment generating function is finite. Choose $r_\star$ with maximal variance and, among ties, maximal mean. For $r\ne r_\star$, either $v_r<v_{r_\star}$ or $v_r=v_{r_\star}$ and $\nu_r<\nu_{r_\star}$. Dividing Equation~\eqref{eq:gaussian_density_linear_independence_mgf} by $\exp\{\nu_{r_\star}t+v_{r_\star}t^2/2\}$ therefore gives
\[
0=a_{r_\star}
+\sum_{\substack{r=1\\r\ne r_\star}}^{q}
a_r\exp\!\left\{(\nu_r-\nu_{r_\star})t
+\frac12(v_r-v_{r_\star})t^2\right\}
\longrightarrow a_{r_\star}
\quad(t\to\infty).
\]
Thus $a_{r_\star}=0$. Repeating the argument for the remaining densities gives $a_r=0$ for every $r$.

The products $f_r(x)f_s(y)$ are also linearly independent. Indeed, if
\[
\sum_{r=1}^{q}\sum_{s=1}^{q}c_{rs}f_r(x)f_s(y)=0
\qquad\text{for almost every }(x,y)\in\mathbb R^2,
\]
Fubini's theorem and linear independence in $x$ give $\sum_{s=1}^{q}c_{rs}f_s(y)=0$ for almost every $y$ and every $r$. Linear independence in $y$ then gives $c_{rs}=0$ for all $r,s$.

For each block $(k,\ell)$, write
\[
f_{k\ell}(x)=\frac{1}{\sigma_{k\ell}}
\phi\!\left(\frac{x-\mu_{k\ell}}{\sigma_{k\ell}}\right),
\qquad k=1,\ldots,K,\quad \ell=1,\ldots,L,
\]
and let $\widetilde f_{k\ell}$ denote the corresponding density under $\widetilde{\boldsymbol\theta}$. By Equations~\eqref{eq:label_prior} and~\eqref{eq:block_gaussian_law_revised}, equality of the one-cell marginals gives
\[
\sum_{k=1}^{K}\sum_{\ell=1}^{L}\pi_k\rho_\ell f_{k\ell}(x)
=\sum_{k=1}^{K}\sum_{\ell=1}^{L}
\widetilde\pi_k\widetilde\rho_\ell\widetilde f_{k\ell}(x)
\qquad\text{for almost every }x\in\mathbb R.
\]
Subtracting the two sides and collecting equal Gaussian densities gives a linear relation. By linear independence, every coefficient vanishes. Since all weights are positive and the block densities are pairwise distinct within each representation, each density occurs exactly once on each side, with the same weight. The matching therefore defines a bijection $h$ of $\{1,\ldots,K\}\times\{1,\ldots,L\}$ such that
\begin{equation}\label{eq:identifiability_block_matching}
f_{k\ell}=\widetilde f_{uv},
\qquad
\pi_k\rho_\ell=\widetilde\pi_u\widetilde\rho_v,
\qquad (u,v)=h(k,\ell).
\end{equation}

Since $n,m\geq2$, consider $(Y_{11},Y_{12})$ and $(Y_{11},Y_{21})$. The label law~\eqref{eq:label_prior} and conditional independence give their respective densities
\begin{equation}\label{eq:identifiability_pair_densities}
\begin{aligned}
g_R(x,y)
&=\sum_{k=1}^{K}\sum_{\ell=1}^{L}\sum_{\ell'=1}^{L}
\pi_k\rho_\ell\rho_{\ell'}f_{k\ell}(x)f_{k\ell'}(y),\\
g_C(x,y)
&=\sum_{\ell=1}^{L}\sum_{k=1}^{K}\sum_{k'=1}^{K}
\rho_\ell\pi_k\pi_{k'}f_{k\ell}(x)f_{k'\ell}(y).
\end{aligned}
\end{equation}
The same expressions with tilded parameters and densities hold under $\widetilde{\boldsymbol\theta}$.

For arbitrary blocks $(k,\ell)$ and $(k',\ell')$, let $(u,v)=h(k,\ell)$ and $(u',v')=h(k',\ell')$. Reindex the tilded expressions in Equation~\eqref{eq:identifiability_pair_densities} using Equation~\eqref{eq:identifiability_block_matching}. Equality of the pair densities and linear independence of the products $f_{k\ell}(x)f_{k'\ell'}(y)$ give
\[
\begin{aligned}
\mathbb I\{k=k'\}\pi_k\rho_\ell\rho_{\ell'}
&=\mathbb I\{u=u'\}\widetilde\pi_u\widetilde\rho_v\widetilde\rho_{v'},\\
\mathbb I\{\ell=\ell'\}\rho_\ell\pi_k\pi_{k'}
&=\mathbb I\{v=v'\}\widetilde\rho_v\widetilde\pi_u\widetilde\pi_{u'}.
\end{aligned}
\]
Positivity of the class proportions implies
\begin{equation}\label{eq:identifiability_sharing_relations}
k=k'\ \text{if and only if}\ u=u',
\qquad
\ell=\ell'\ \text{if and only if}\ v=v'.
\end{equation}

By Equation~\eqref{eq:identifiability_sharing_relations}, the first coordinate of $h(k,\ell)$ is constant over $\ell$ for each $k$, and the second is constant over $k$ for each $\ell$. Denote these values by $s(k)$ and $t(\ell)$, respectively. Then $h(k,\ell)=(s(k),t(\ell))$. Applying the reverse implications in Equation~\eqref{eq:identifiability_sharing_relations} to $(k,1),(k',1)$ and $(1,\ell),(1,\ell')$ gives
\[
s(k)=s(k')\ \Longrightarrow\ k=k',
\qquad
t(\ell)=t(\ell')\ \Longrightarrow\ \ell=\ell'.
\]
Thus $s$ and $t$ are injective maps of the respective finite label sets to themselves, and hence are permutations.

Equal Gaussian densities have equal means and variances, so Equation~\eqref{eq:identifiability_block_matching} now gives
\[
\begin{aligned}
(\mu_{k\ell},\sigma^2_{k\ell})
&=(\widetilde\mu_{s(k),t(\ell)},\widetilde\sigma^2_{s(k),t(\ell)}),\\
\pi_k\rho_\ell&=\widetilde\pi_{s(k)}\widetilde\rho_{t(\ell)}.
\end{aligned}
\]
Summing the weight identity and using the normalization of the class proportions and the permutations $s,t$ yields
\[
\begin{aligned}
\pi_k
&=\sum_{\ell=1}^{L}\pi_k\rho_\ell
=\widetilde\pi_{s(k)}\sum_{\ell=1}^{L}\widetilde\rho_{t(\ell)}
=\widetilde\pi_{s(k)},\\
\rho_\ell
&=\sum_{k=1}^{K}\pi_k\rho_\ell
=\widetilde\rho_{t(\ell)}\sum_{k=1}^{K}\widetilde\pi_{s(k)}
=\widetilde\rho_{t(\ell)}.
\end{aligned}
\]
Therefore $\boldsymbol\theta\sim\widetilde{\boldsymbol\theta}$.

Conversely, let $\boldsymbol\theta\sim\widetilde{\boldsymbol\theta}$ with permutations $s,t$. For each label configuration, set $z_i'=s(z_i)$ and $w_j'=t(w_j)$. Equations~\eqref{eq:label_prior} and~\eqref{eq:fixed_label_observed_likelihood} give
\[
p_{\boldsymbol\theta}(\mathbf z,\mathbf w)
p_{\boldsymbol\theta}(\mathbf d\mid\mathbf z,\mathbf w)
=
p_{\widetilde{\boldsymbol\theta}}(\mathbf z',\mathbf w')
p_{\widetilde{\boldsymbol\theta}}(\mathbf d\mid\mathbf z',\mathbf w').
\]
Since this relabeling is bijective and the identity holds for every possible report matrix $\mathbf d$, summing over the label configurations in Equation~\eqref{eq:observed_data_likelihood_revised} gives $\mathbb P_{\boldsymbol\theta}^{\mathbf D}=\mathbb P_{\widetilde{\boldsymbol\theta}}^{\mathbf D}$.
\end{proof}

\subsection{Proof of Lemma~\ref{lem:left_censor_fisher_information}}
\label{app:proof_left_censor_fisher_information}

\begin{proof}
First suppose both reporting limits are finite.
Let $Z=(Y-\mu)/\sigma$, with $\alpha,\beta,p_L,p_R,q$ as defined
in Lemma~\ref{lem:left_censor_fisher_information}.
Differentiating $\log p_L$, $-\log\sigma+\log\phi(Z)$, and
$\log p_R$ in the three reporting regions gives the score for
$(\mu,\sigma^2)$:
\begin{equation}\label{eq:lower_upper_report_score}
U=
\begin{cases}
-\dfrac{\phi(\alpha)}{p_L}
\left(
\dfrac{1}{\sigma},
\dfrac{\alpha}{2\sigma^2}
\right)^\top,
& Z\leq\alpha,\\[6pt]
\left(
\dfrac{Z}{\sigma},
\dfrac{Z^2-1}{2\sigma^2}
\right)^\top,
& \alpha<Z<\beta,\\[6pt]
\dfrac{\phi(\beta)}{p_R}
\left(
\dfrac{1}{\sigma},
\dfrac{\beta}{2\sigma^2}
\right)^\top,
& Z\geq\beta.
\end{cases}
\end{equation}

Taking the expected outer product of the score in
Equation~\eqref{eq:lower_upper_report_score} gives
Equation~\eqref{eq:lower_upper_fisher_information}, with
\[
\begin{aligned}
A(\alpha,\beta)
&=
\int_\alpha^\beta z^2\phi(z)\,dz
+\frac{\phi(\alpha)^2}{p_L}
+\frac{\phi(\beta)^2}{p_R},\\[4pt]
B(\alpha,\beta)
&=
\int_\alpha^\beta z(z^2-1)\phi(z)\,dz
+\frac{\alpha\phi(\alpha)^2}{p_L}
+\frac{\beta\phi(\beta)^2}{p_R},\\[4pt]
C(\alpha,\beta)
&=
\int_\alpha^\beta (z^2-1)^2\phi(z)\,dz
+\frac{\alpha^2\phi(\alpha)^2}{p_L}
+\frac{\beta^2\phi(\beta)^2}{p_R}.
\end{aligned}
\]

Integration by parts with $\phi'(z)=-z\phi(z)$ gives
\begin{equation}\label{eq:gaussian_tail_moment_recurrence}
\int_\alpha^\infty z^s\phi(z)\,dz
=
\alpha^{s-1}\phi(\alpha)
+
(s-1)\int_\alpha^\infty z^{s-2}\phi(z)\,dz,
\qquad s=2,3,4.
\end{equation}
The integrals over $(\alpha,\beta)$ follow by subtracting the
corresponding tail integrals.
Together with
$\int_\alpha^\beta\phi(z)\,dz=q$ and
$\int_\alpha^\beta z\phi(z)\,dz=\phi(\alpha)-\phi(\beta)$,
this gives the coefficient expressions in
Lemma~\ref{lem:left_censor_fisher_information}.

The formulas extend to $c_L=-\infty$ or $c_U=\infty$ by taking limits.
The corresponding tail terms vanish, giving the information matrices
for one-sided censoring.
When both limits are infinite, the matrix reduces to
$\operatorname{diag}\{\sigma^{-2},(2\sigma^4)^{-1}\}$.

To prove positive definiteness, suppose $a=(a_1,a_2)^\top$ satisfies
$a^\top\mathcal I a=0$.
Then
\[
0=\E\{(a^\top U)^2\}
\geq
\int_\alpha^\beta
\left\{
\frac{a_1z}{\sigma}
+
\frac{a_2(z^2-1)}{2\sigma^2}
\right\}^2
\phi(z)\,dz
\geq0.
\]
Since $\phi(z)>0$, the polynomial in braces vanishes almost everywhere
on the nonempty open interval $(\alpha,\beta)$.
It therefore vanishes identically.
Its coefficients give $a_1=a_2=0$, so $\mathcal I$ is positive definite.
The same argument applies when either endpoint is infinite.

For left censoring at a finite threshold $c$, set $c_L=c$ and
$c_U=\infty$, and write $p=\Phi(\alpha)$ and $r=\phi(\alpha)/p$.
The coefficients become
\begin{equation}\label{eq:left_censor_information_functions}
\begin{aligned}
A(\alpha,\infty)
&=1-p+\phi(\alpha)\{\alpha+r\},\\
B(\alpha,\infty)
&=\phi(\alpha)\{\alpha^2+1+\alpha r\},\\
C(\alpha,\infty)
&=2(1-p)+\phi(\alpha)\{\alpha^3+\alpha+\alpha^2r\}.
\end{aligned}
\end{equation}

For $c'>c$, the report at $c'$ is a parameter-independent measurable
function of the report at $c$.
Fisher information cannot increase under this transformation.
Since $p$ increases with $c$ for fixed $(\mu,\sigma^2)$, the stated
Loewner monotonicity follows.

Finally, hold $(\mu,\sigma^2)$ fixed.
As $\alpha\to-\infty$, $p/\{\phi(\alpha)/(-\alpha)\}\to1$, so
$|\alpha|^s\phi(\alpha)^2/p\to0$ for $s=0,1,2$.
These terms also vanish as $\alpha\to\infty$, since $p\to1$.
Together with the vanishing polynomial multiples of $\phi(\alpha)$,
Equation~\eqref{eq:left_censor_information_functions} gives
\[
\lim_{p\to0}\mathcal I(\mu,\sigma^2;c,\infty)
=
\begin{pmatrix}
\sigma^{-2} & 0\\[2pt]
0 & (2\sigma^4)^{-1}
\end{pmatrix},
\qquad
\lim_{p\to1}\mathcal I(\mu,\sigma^2;c,\infty)=\mathbf0.
\]
\end{proof}

\subsection{Information from a censoring indicator}\label{app:censoring_indicator_information}

For a finite left-censoring threshold $c$, write
$\alpha=(c-\mu)/\sigma$, $p=\Phi(\alpha)$, and
$r=\phi(\alpha)/p$.
Suppose only $C=\ind\{Y\leq c\}\sim\operatorname{Bernoulli}(p)$ is observed. Its score for $(\mu,\sigma^2)$ is $(C-p)\nabla p/\{p(1-p)\}$, where $\nabla p=-\phi(\alpha)(\sigma^{-1},\alpha/(2\sigma^2))^\top$. Since $E\{(C-p)^2\}=p(1-p)$,
\[
\mathcal I_{\mathrm{ind}}(\mu,\sigma^2;c)
=\frac{(\nabla p)(\nabla p)^\top}{p(1-p)}
=\frac{\phi(\alpha)^2}{p(1-p)}
\begin{pmatrix}
\sigma^{-2} & \alpha/(2\sigma^3)\\[3pt]
\alpha/(2\sigma^3) & \alpha^2/(4\sigma^4)
\end{pmatrix}.
\]
The indicator law depends on $(\mu,\sigma^2)$ only through $\alpha$, and $\nabla p\neq0$ implies $\operatorname{rank}(\mathcal I_{\mathrm{ind}})=1$. In contrast, the full report has positive definite information by
the left-censoring special case of
Lemma~\ref{lem:left_censor_fisher_information}.

With the other parameter known, division by the corresponding exact Gaussian information gives
\[
R_\mu(\alpha)=\frac{\phi(\alpha)^2}{p(1-p)},
\qquad
R_{\sigma^2}(\alpha)=\frac{\alpha^2\phi(\alpha)^2}{2p(1-p)}
=\frac{\alpha^2}{2}R_\mu(\alpha).
\]
Both ratios are even and tend to zero as $|\alpha|\to\infty$. Differentiation gives
\[
\frac{d}{d\alpha}\log R_\mu(\alpha)
=-2\alpha+\frac{\phi(\alpha)(2p-1)}{p(1-p)}.
\]
For $Z=(Y-\mu)/\sigma$, a second differentiation and the truncated moments obtained from Equation~\eqref{eq:gaussian_tail_moment_recurrence} give
\[
\begin{aligned}
\frac{d^2}{d\alpha^2}\log R_\mu(\alpha)
&=-(1-\alpha r-r^2)
-\left\{1+\frac{\alpha\phi(\alpha)}{1-p}
-\frac{\phi(\alpha)^2}{(1-p)^2}\right\}\\
&=-\operatorname{Var}(Z\mid Z\leq\alpha)
-\operatorname{Var}(Z\mid Z>\alpha)<0.
\end{aligned}
\]
Thus $\log R_\mu$ is strictly concave, and evenness gives the unique maximizer $\alpha=0$, with
\[
\max_{\alpha\in\mathbb R}R_\mu(\alpha)=R_\mu(0)=\frac{2}{\pi}.
\]

For $\alpha>0$,
\[
\frac{d^2}{d\alpha^2}\log R_{\sigma^2}(\alpha)
=-\frac{2}{\alpha^2}+\frac{d^2}{d\alpha^2}\log R_\mu(\alpha)<0.
\]
Since $R_{\sigma^2}(0)=\lim_{\alpha\to\infty}R_{\sigma^2}(\alpha)=0$ and $R_{\sigma^2}(\alpha)>0$ for $\alpha>0$, its unique positive maximizer satisfies
\begin{equation}\label{eq:censoring_indicator_variance_stationarity}
\frac{d}{d\alpha}\log R_{\sigma^2}(\alpha)
=\frac{2}{\alpha}-2\alpha
+\frac{\phi(\alpha)(2p-1)}{p(1-p)}=0.
\end{equation}
Solving Equation~\eqref{eq:censoring_indicator_variance_stationarity} numerically and using symmetry gives the maximizers $\alpha\approx\pm1.575036263$ and
\[
\max_{\alpha\in\mathbb R}R_{\sigma^2}(\alpha)\approx0.304208966.
\]

\subsection{Proof of Proposition~\ref{prop:weighted_censored_gaussian_update}}\label{app:proof_weighted_censored_gaussian_update}

\begin{proof}
For fixed memberships, substituting the Gaussian density from Equation~\eqref{eq:block_gaussian_law_revised} and the conditional moments into Equation~\eqref{eq:weighted_Q_definition_new} gives, up to an additive constant independent of the Gaussian block parameters,
\begin{equation}\label{eq:weighted_Q_function_new}
Q_t(\boldsymbol\theta)
=-\frac12
\sum_{i=1}^n\sum_{j=1}^m\sum_{k=1}^K\sum_{\ell=1}^L
\gamma_{ij,k\ell}
\left[
\log\sigma_{k\ell}^2
+\frac{
\overline{y^2}_{ij,k\ell}^{(t)}
-2\mu_{k\ell}\overline y_{ij,k\ell}^{(t)}
+\mu_{k\ell}^2
}{\sigma_{k\ell}^2}
\right].
\end{equation}
For each block, differentiation gives
\[
\begin{aligned}
\frac{\partial Q_t}{\partial\mu_{k\ell}}
&=\frac{1}{\sigma_{k\ell}^2}
\left(
\sum_{i=1}^n\sum_{j=1}^m
\gamma_{ij,k\ell}\overline y_{ij,k\ell}^{(t)}
-N_{k\ell}\mu_{k\ell}
\right),\\
\frac{\partial^2 Q_t}{\partial\mu_{k\ell}^2}
&=-\frac{N_{k\ell}}{\sigma_{k\ell}^2}<0.
\end{aligned}
\]
Setting $\partial Q_t/\partial\mu_{k\ell}=0$ gives Equation~\eqref{eq:weighted_mean_update_new}, the unique maximizing mean for every $\sigma_{k\ell}^2>0$. At this mean,
\[
\begin{aligned}
\frac{\partial Q_t}{\partial\sigma_{k\ell}^2}
&=\frac{N_{k\ell}}{2(\sigma_{k\ell}^2)^2}
\left\{
\frac{1}{N_{k\ell}}
\sum_{i=1}^n\sum_{j=1}^m
\gamma_{ij,k\ell}
\left[
\overline{y^2}_{ij,k\ell}^{(t)}
-2\mu_{k\ell}^{(t+1)}\overline y_{ij,k\ell}^{(t)}
+(\mu_{k\ell}^{(t+1)})^2
\right]
-\sigma_{k\ell}^2
\right\}.
\end{aligned}
\]
The weighted average of conditional expected squared deviations is nonnegative. If this average is positive, the derivative is positive below it and negative above it. If it is zero, the derivative is negative for every $\sigma_{k\ell}^2>0$. The maximizer under $\sigma_{k\ell}^2\geq\sigma_{\min}^2$ is thus Equation~\eqref{eq:weighted_variance_update_new}. Since Equation~\eqref{eq:weighted_Q_function_new} separates over blocks, these updates maximize $Q_t$ over the constrained Gaussian block parameter space.

For a censored report, the conditional density is the Gaussian density restricted to the reported interval and divided by $\lambda_{ij,k\ell}(d_{ij};\boldsymbol\theta^{(t)})$. For an exact report, the conditional expectation is evaluation at $d_{ij}$. Hence, for $\gamma_{ij,k\ell}>0$ and feasible $\boldsymbol\theta$, Jensen's inequality gives
\begin{equation}\label{eq:weighted_report_jensen_new}
\begin{aligned}
&\log\frac{\lambda_{ij,k\ell}(d_{ij};\boldsymbol\theta)}
{\lambda_{ij,k\ell}(d_{ij};\boldsymbol\theta^{(t)})}\\
&\quad=\log\E_{\boldsymbol\theta^{(t)}}
\left[
\frac{p_{\boldsymbol\theta}(Y_{ij}\mid Z_i=k,W_j=\ell)}
{p_{\boldsymbol\theta^{(t)}}(Y_{ij}\mid Z_i=k,W_j=\ell)}
\mathrel{\Big|}D_{ij}=d_{ij},Z_i=k,W_j=\ell
\right]\\
&\quad\geq\E_{\boldsymbol\theta^{(t)}}
\left[
\log\frac{p_{\boldsymbol\theta}(Y_{ij}\mid Z_i=k,W_j=\ell)}
{p_{\boldsymbol\theta^{(t)}}(Y_{ij}\mid Z_i=k,W_j=\ell)}
\mathrel{\Big|}D_{ij}=d_{ij},Z_i=k,W_j=\ell
\right].
\end{aligned}
\end{equation}
Taking $\boldsymbol\theta=\boldsymbol\theta^{(t+1)}$ in Equation~\eqref{eq:weighted_report_jensen_new} and summing with weights $\gamma_{ij,k\ell}$ gives, by Equation~\eqref{eq:weighted_Q_definition_new} and the constrained maximization of $Q_t$,
\[
\begin{aligned}
&\sum_{i=1}^n\sum_{j=1}^m\sum_{k=1}^K\sum_{\ell=1}^L
\gamma_{ij,k\ell}
\log\frac{\lambda_{ij,k\ell}(d_{ij};\boldsymbol\theta^{(t+1)})}
{\lambda_{ij,k\ell}(d_{ij};\boldsymbol\theta^{(t)})}\\
&\quad\geq Q_t(\boldsymbol\theta^{(t+1)})
-Q_t(\boldsymbol\theta^{(t)})
\geq0,
\end{aligned}
\]
where the last inequality uses feasibility of $\boldsymbol\theta^{(t)}$. This proves Equation~\eqref{eq:block_update_monotonicity}.
\end{proof}

\subsection{Proof of Lemma~\ref{lem:complete_left_censoring}}\label{app:proof_complete_left_censoring}

\begin{proof}
For every finite feasible pair $(\mu_{k\ell},\sigma_{k\ell})$ and every positive-weight report, the finite threshold $b_{ij}$ gives $0<\Phi\{(b_{ij}-\mu_{k\ell})/\sigma_{k\ell}\}<1$. Since $N_{k\ell}>0$,
\[
\sum_{i=1}^n\sum_{j=1}^m
\gamma_{ij,k\ell}
\log\Phi\!\left(\frac{b_{ij}-\mu_{k\ell}}{\sigma_{k\ell}}\right)<0.
\]
Fix $\sigma_{k\ell}=\sigma_{\min}$. For every positive-weight report, $(b_{ij}-\mu_{k\ell})/\sigma_{\min}\to\infty$ as $\mu_{k\ell}\to-\infty$. The sum is finite, so
\[
\lim_{\mu_{k\ell}\to-\infty}
\sum_{i=1}^n\sum_{j=1}^m
\gamma_{ij,k\ell}
\log\Phi\!\left(\frac{b_{ij}-\mu_{k\ell}}{\sigma_{\min}}\right)=0.
\]
Thus the supremum is zero and is not attained at a finite parameter pair. The right-censoring result follows by applying the same argument to $-Y_{ij}$.
\end{proof}

\subsection{Derivation of the BIC approximation to the ICL}\label{app:icl_bic_derivation}

Fix $M=(K,L)$ and hold the MAP labels $\widehat{\mathbf z}_M,\widehat{\mathbf w}_M$ fixed. Let
\[
n_k=\sum_{i=1}^n\mathbb I\{\widehat z_{i,M}=k\},
\qquad
m_\ell=\sum_{j=1}^m\mathbb I\{\widehat w_{j,M}=\ell\}.
\]
Take independent priors for the three parameter groups, so that
$\nu_M(\boldsymbol\theta)=\nu_\pi(\boldsymbol\pi)\nu_\rho(\boldsymbol\rho)\nu_G(\boldsymbol\mu,\boldsymbol\sigma^2)$.
By Equations~\eqref{eq:label_prior} and~\eqref{eq:fixed_label_observed_likelihood},
\[
p_{\boldsymbol\theta}(\mathbf d,\mathbf z,\mathbf w)
=
p_{\boldsymbol\pi}(\mathbf z)
p_{\boldsymbol\rho}(\mathbf w)
p_{\boldsymbol\mu,\boldsymbol\sigma^2}(\mathbf d\mid\mathbf z,\mathbf w),
\]
and Equation~\eqref{eq:integrated_classification_likelihood} becomes
\[
\begin{aligned}
\operatorname{ICL}(M)
&=\log\int p_{\boldsymbol\pi}(\widehat{\mathbf z}_M)
\nu_\pi(\boldsymbol\pi)\,d\boldsymbol\pi
+\log\int p_{\boldsymbol\rho}(\widehat{\mathbf w}_M)
\nu_\rho(\boldsymbol\rho)\,d\boldsymbol\rho\\
&\quad+\log\int
p_{\boldsymbol\mu,\boldsymbol\sigma^2}
(\mathbf d\mid\widehat{\mathbf z}_M,\widehat{\mathbf w}_M)
\nu_G(\boldsymbol\mu,\boldsymbol\sigma^2)
\,d\boldsymbol\mu\,d\boldsymbol\sigma^2.
\end{aligned}
\]
Let $\widetilde{\boldsymbol\theta}_M$ be the maximizer of the classification likelihood over $\Theta_M$.

As $n,m\to\infty$, assume $K,L$ remain fixed and all $n_k/n,m_\ell/m$ converge to positive limits. Assume isolated interior likelihood maximizers converging to interior points, with limiting block variances strictly above $\sigma_{\min}^2$. The prior factors $\nu_\pi,\nu_\rho,\nu_G$ are fixed, proper, smooth, and positive near these limits.

The row-proportion log likelihood is $\sum_{k=1}^K n_k\log\pi_k$, with maximizer $\widetilde\pi_{k,M}=n_k/n$. For $K>1$, using $\pi_1,\ldots,\pi_{K-1}$ as free coordinates and $\pi_K=1-\sum_{k=1}^{K-1}\pi_k$ gives the negative Hessian
\[
(H_\pi)_{ab}
=n\left\{\frac{\mathbb I\{a=b\}}{\widetilde\pi_{a,M}}
+\frac{1}{\widetilde\pi_{K,M}}\right\},
\qquad a,b=1,\ldots,K-1.
\]
The matrix determinant lemma and the analogous column calculation, with $\widetilde\rho_{\ell,M}=m_\ell/m$, give
\begin{equation}\label{eq:icl_proportion_hessian_determinants}
\det H_\pi
=\frac{n^{K-1}}{\prod_{k=1}^K(n_k/n)},
\qquad
\det H_\rho
=\frac{m^{L-1}}{\prod_{\ell=1}^L(m_\ell/m)}.
\end{equation}
When $K=1$ or $L=1$, the corresponding proportion is fixed at one, its integral equals one, and its Hessian determinant is interpreted as one.

For each Gaussian block, define
\[
\ell_{k\ell}(\mu_{k\ell},\sigma_{k\ell}^2)
=\sum_{i:\,\widehat z_{i,M}=k}
\sum_{j:\,\widehat w_{j,M}=\ell}
\log\lambda_{ij,k\ell}(d_{ij};\boldsymbol\theta).
\]
Let $H_{k\ell}$ be its $2\times2$ negative Hessian at the maximizing block parameters. Assume
\[
\frac{H_{k\ell}}{n_km_\ell}\longrightarrow I_{k\ell}\succ0,
\qquad k=1,\ldots,K,\quad\ell=1,\ldots,L,
\]
with each $I_{k\ell}$ finite. Since the Gaussian log likelihood separates over blocks, its Hessian is block diagonal and
\begin{equation}\label{eq:icl_gaussian_hessian_order}
\begin{aligned}
\sum_{k=1}^K\sum_{\ell=1}^L\log\det H_{k\ell}
&=2\sum_{k=1}^K\sum_{\ell=1}^L\log(n_km_\ell)
+\sum_{k=1}^K\sum_{\ell=1}^L
\log\det\!\left(\frac{H_{k\ell}}{n_km_\ell}\right)\\
&=2KL\log(nm)+O(1).
\end{aligned}
\end{equation}
The last equality uses $\log(n_km_\ell)=\log(nm)+\log(n_k/n)+\log(m_\ell/m)$ together with the positive limiting class fractions and $I_{k\ell}\succ0$. Known reporting limits change the cell information, as in Lemma~\ref{lem:left_censor_fisher_information}, but add no parameters. The positive definite block limits are required also when reporting limits vary across cells.

For the row, column, and Gaussian likelihood factors at scales $N=n,m,nm$, respectively, assume the quadratic remainder is uniformly vanishing on a ball of radius $r_N$ around the maximizer, with $r_N\to0$ and $\sqrt N r_N\to\infty$. Assume also that the prior-weighted integral outside this ball is negligible relative to the local integral. Integrating these local quadratic approximations in Equation~\eqref{eq:integrated_classification_likelihood}, with $K-1$, $L-1$, and $2KL$ free parameters, gives
\begin{equation}\label{eq:icl_laplace_expansion_appendix}
\begin{aligned}
\operatorname{ICL}(M)
&=\log p_{\widetilde{\boldsymbol\theta}_M}
(\mathbf d,\widehat{\mathbf z}_M,\widehat{\mathbf w}_M)
+\log\nu_M(\widetilde{\boldsymbol\theta}_M)\\
&\quad+\frac{K+L-2+2KL}{2}\log(2\pi)
-\frac12\log\det H_\pi-\frac12\log\det H_\rho\\
&\quad-\frac12\sum_{k=1}^K\sum_{\ell=1}^L\log\det H_{k\ell}+o(1).
\end{aligned}
\end{equation}
The prior term is bounded by the assumed convergence and positivity. Substituting Equations~\eqref{eq:icl_proportion_hessian_determinants} and~\eqref{eq:icl_gaussian_hessian_order} into Equation~\eqref{eq:icl_laplace_expansion_appendix} therefore yields
\begin{equation}\label{eq:icl_bic_expansion_appendix}
\begin{aligned}
\operatorname{ICL}(M)
&=\max_{\boldsymbol\theta\in\Theta_M}
\log p_{\boldsymbol\theta}
(\mathbf d,\widehat{\mathbf z}_M,\widehat{\mathbf w}_M)\\
&\quad-\frac{K-1}{2}\log n
-\frac{L-1}{2}\log m
-\frac{2KL}{2}\log(nm)+O(1).
\end{aligned}
\end{equation}
The bounded terms may depend on $M$, the priors, the limiting class fractions, and the censoring design. Omitting them from Equation~\eqref{eq:icl_bic_expansion_appendix} and replacing the maximized log likelihood by the plug-in evaluation defined in Section~\ref{subsec:selection_row_column_class_numbers} gives Equation~\eqref{eq:censored_icl_bic}.

\clearpage

\section{Additional simulation details and results}\label{app:simulation_metrics}
\setcounter{equation}{0}
\renewcommand{\theequation}{B\arabic{equation}}
\renewcommand{\theHequation}{B\arabic{equation}}

\subsection{Common simulation details}\label{app:simulation_common_details}

\paragraph{Censoring calibration.}
For each positive-censoring setting, 100 calibration matrices were generated independently of the evaluation matrices and pooled to define \(\widehat F_{\mathrm{cal}}\) in Equation~\eqref{eq:simulation_detection_limits}.

\paragraph{Numerical settings.}
Each fit requested 20 valid starts from at most 200 attempts and used at most 400 iterations, a convergence tolerance of \(10^{-7}\), and a variance floor of \(10^{-8}\). Effective row class, column class, and block masses at or below \(10^{-8}\) were treated as zero.

\paragraph{Performance measures.}
For two partitions \(\mathbf u=(u_1,\ldots,u_N)^\intercal\) and \(\mathbf v=(v_1,\ldots,v_N)^\intercal\), define
\(n_{ab}=\sum_{r=1}^{N}\ind\{u_r=a,v_r=b\}\),
\(n_{a\cdot}=\sum_b n_{ab}\), and
\(n_{\cdot b}=\sum_a n_{ab}\).
The adjusted Rand index \citep{hubert1985} is
\begin{equation}\label{eq:simulation_ari}
\operatorname{ARI}(\mathbf u,\mathbf v)
=
\frac{
\displaystyle
\sum_{a,b}\binom{n_{ab}}{2}
-
\frac{
\displaystyle
\sum_a\binom{n_{a\cdot}}{2}
\sum_b\binom{n_{\cdot b}}{2}
}{
\displaystyle
\binom{N}{2}
}
}{
\displaystyle
\frac12
\left\{
\sum_a\binom{n_{a\cdot}}{2}
+
\sum_b\binom{n_{\cdot b}}{2}
\right\}
-
\frac{
\displaystyle
\sum_a\binom{n_{a\cdot}}{2}
\sum_b\binom{n_{\cdot b}}{2}
}{
\displaystyle
\binom{N}{2}
}
}.
\end{equation}
The marginal recovery measures were
\[
\operatorname{Row\ ARI}=\operatorname{ARI}(\mathbf z^\star,\widehat{\mathbf z}),
\qquad
\operatorname{Column\ ARI}=\operatorname{ARI}(\mathbf w^\star,\widehat{\mathbf w}).
\]
Block mean estimation was assessed after separate row and column label alignment by
\[
\operatorname{RMSE}_{\mu}
=
\min_{p,q}
\left[
\sum_{k=1}^{K}
\sum_{\ell=1}^{L}
\frac{n_km_\ell}{nm}
\left\{
\widehat\mu_{p(k),q(\ell)}
-
\mu_{k\ell}^\star
\right\}^{2}
\right]^{1/2},
\]
where \(p\) and \(q\) range over permutations of the row and column class labels.

After the same separate row and column label alignment, let
$\widehat{\mu}^{\,\mathrm{al}}_{k\ell}$ denote the aligned fitted mean for
block $(k,\ell)$. Let $(k_{\max},\ell_{\max})$ and
$(k_{\min},\ell_{\min})$ denote the blocks with the largest and smallest true
means, respectively. The block mean contrast ratio was
\[
\frac{
\widehat{\mu}^{\,\mathrm{al}}_{k_{\max}\ell_{\max}}
-
\widehat{\mu}^{\,\mathrm{al}}_{k_{\min}\ell_{\min}}
}{
\mu^\star_{k_{\max}\ell_{\max}}
-
\mu^\star_{k_{\min}\ell_{\min}}
}.
\]
A value below one means that the estimated contrast is smaller than the true
contrast, whereas a value above one means that it is larger.

\paragraph{Monte Carlo summaries.}
Within each signal strength, summaries across censoring conditions were first computed for each latent matrix and then averaged across matrices. Means and paired contrasts were computed within each Monte Carlo unit, and two-sided 95\% Monte Carlo confidence intervals used \(t\) intervals over the resulting unit-level summaries. For the class-size comparison, the balanced reference and the three imbalanced structures formed one matched Monte Carlo unit.

\subsection{Experiment I matching and implementation details}\label{app:experiment_one_design}

\paragraph{Matched designs.}
For each signal strength and replicate in the balanced factorial, the same latent matrix was reported under all nine combinations of censoring rate and direction. In the class-size extension, the three class sizes were assigned cyclically to the row and column signal profiles across replicates. The same assignment was used in both dimensions under two-way imbalance. Within each replicate, the balanced reference and the three imbalanced structures used common random numbers, with a separate calibration sample for each class-size structure. In the high-censoring comparison, the same latent matrix was reported at target censoring rates of 30\%, 50\%, and 75\%, with independently calibrated limits at each rate. The 30\% condition was the corresponding balanced-factorial condition.

\paragraph{Completion procedures.}
On the analysis scale, Boundary substitution replaced left- and right-censored reports by \(c_L\) and \(c_U\). Half or double substitution used \(c_L-\log 2\) and \(c_U+\log 2\). Truncated normal imputation estimated one Gaussian distribution from the exact reports and drew each censored value from that distribution restricted to its reported interval.

\paragraph{Multiple imputation.}
For each column, the sampler used
\[
\sigma_j^2\sim\operatorname{Inv\text{-}Gamma}(\alpha_0,\beta_0),
\qquad
\mu_j\mid\sigma_j^2\sim\normal\!\left(\mu_0,\frac{\sigma_j^2}{\kappa_0}\right),
\]
where the inverse-gamma density is proportional to
\((\sigma_j^2)^{-\alpha_0-1}\exp(-\beta_0/\sigma_j^2)\).
The hyperparameters were
\[
\mu_0=\overline d_{\mathrm{exact}},
\qquad
s_0^2=\operatorname{Var}(d_{\mathrm{exact}}),
\qquad
\kappa_0=0.01,
\qquad
\alpha_0=2.01,
\qquad
\beta_0=1.01s_0^2.
\]
Here $d_{\mathrm{exact}}$ collects the exact reported values.
When fewer than two exact reports were available, \(\mu_0\) and \(s_0^2\) were computed after replacing left-censored reports by their upper limits, right-censored reports by their lower limits, and interval-censored reports by their interval midpoints. Each iteration drew column-specific means and variances from the conjugate posterior and then drew censored entries from the corresponding truncated normal distributions. Exact reports remained fixed. The sampler discarded 50 iterations and retained ten completed matrices 20 iterations apart. The final partition was the fitted partition with the largest mean pairwise CARI among the ten fits. Fitted block parameters were aligned to that partition before averaging.

\paragraph{Matched initializations.}
All matched procedures used the same ordered pool of candidate row and column partitions, and the ten multiple-imputation fits reused this pool. Runs were accepted or discarded under the criteria in Section~\ref{subsec:unified_membership_gem}.

\subsection{Additional results for Experiment I}\label{app:experiment_one_additional_results}

\begingroup
\renewcommand{\thefigure}{B1}
\renewcommand{\theHfigure}{B1}
\begin{figure}[H]
\centering
\includegraphics[width=0.99\textwidth]{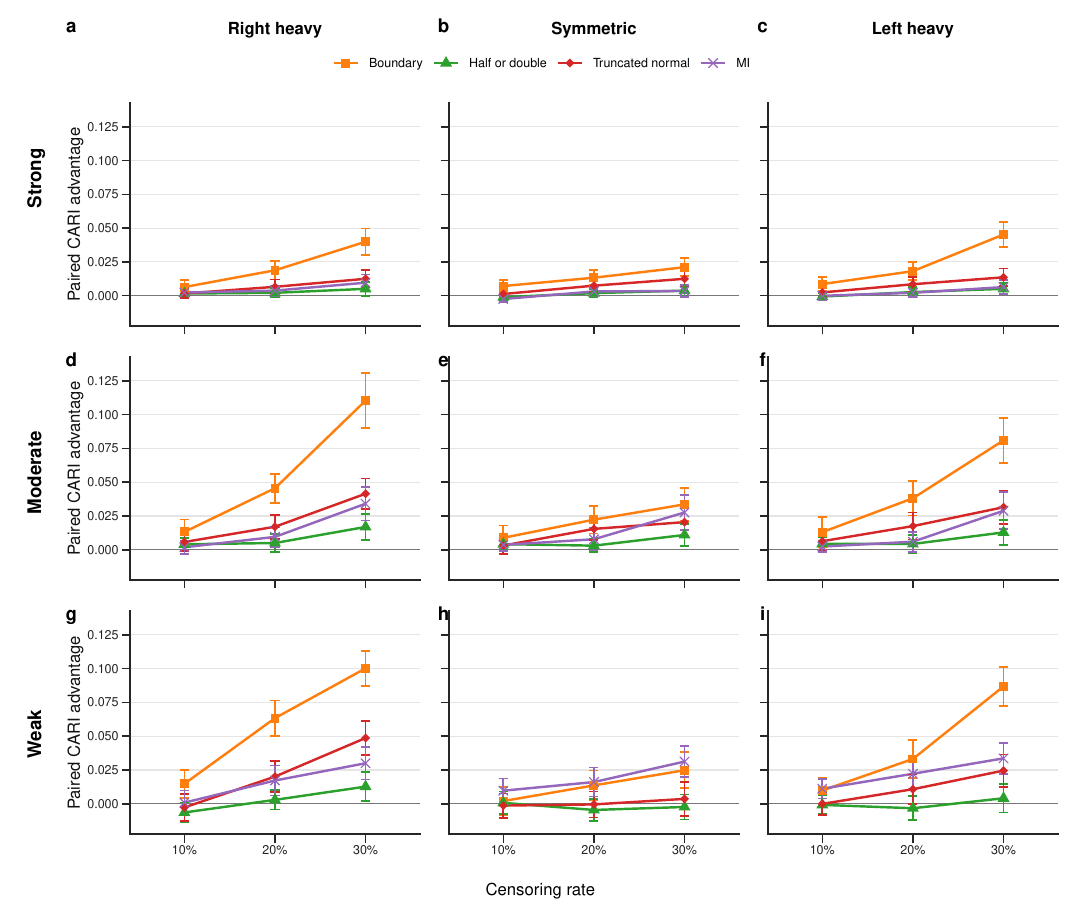}
\refstepcounter{figure}\label{figfullfactorialcomparison}
\par\smallskip
\begin{minipage}{0.98\textwidth}
\small\textbf{Figure \thefigure.} Complete paired CARI comparisons under \(\mathsf{SS}\) in the balanced factorial design. Rows correspond to strong, moderate, and weak signal, and columns correspond to right heavy, symmetric, and left heavy censoring. Each panel shows the paired CARI advantages of CGLBM over the four completion procedures across censoring rates. Positive values favor CGLBM. Error bars show 95\% Monte Carlo confidence intervals.
\end{minipage}
\end{figure}
\endgroup
\addtocounter{figure}{-1}

\clearpage
\begingroup
\renewcommand{\thefigure}{B2}
\renewcommand{\theHfigure}{B2}
\begin{figure}[H]
\centering
\includegraphics[width=0.98\textwidth]{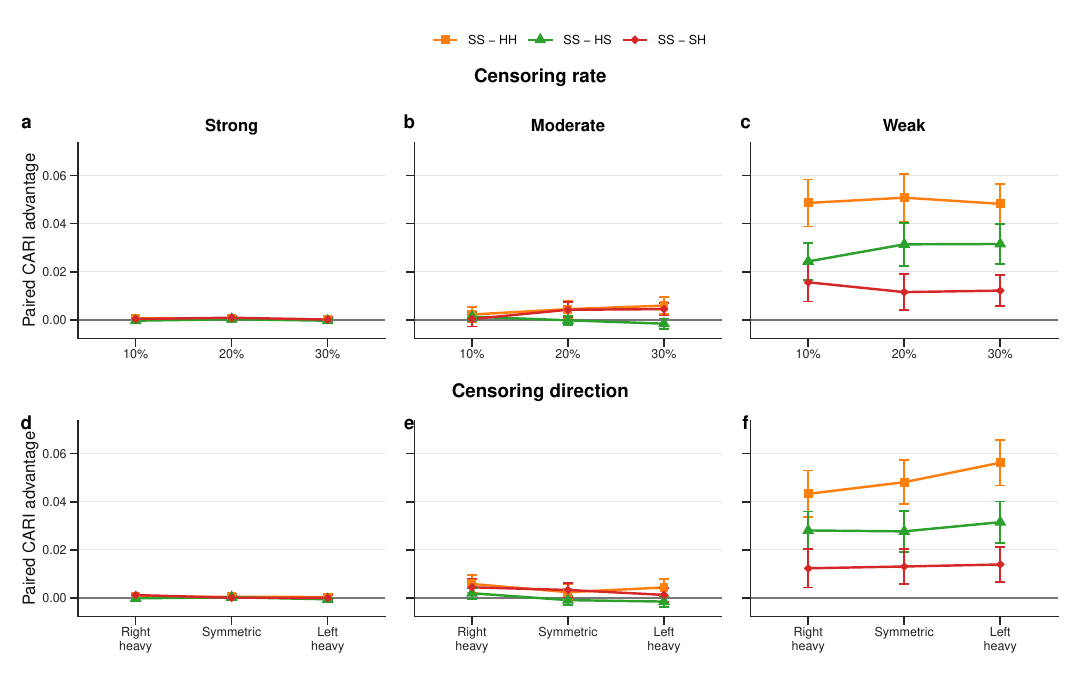}
\refstepcounter{figure}\label{figsimulationconfiguration}
\par\smallskip
\begin{minipage}{0.98\textwidth}
\small\textbf{Figure \thefigure.} Paired CARI advantages of \(\mathsf{SS}\) over \(\mathsf{HH}\), \(\mathsf{HS}\), and \(\mathsf{SH}\). Figures~\ref{figsimulationconfiguration}(a)--\ref{figsimulationconfiguration}(c) show advantages by censoring rate for strong, moderate, and weak signal, respectively, averaging over censoring directions within each signal strength. Figures~\ref{figsimulationconfiguration}(d)--\ref{figsimulationconfiguration}(f) show advantages by censoring direction for the same signal strengths, averaging over censoring rates within each signal strength. Positive values favor \(\mathsf{SS}\). Error bars show 95\% Monte Carlo confidence intervals.
\end{minipage}
\end{figure}
\endgroup
\addtocounter{figure}{-1}

\clearpage
\begingroup
\renewcommand{\thefigure}{B3}
\renewcommand{\theHfigure}{B3}
\begin{figure}[H]
\centering
\includegraphics[width=0.80\textwidth]{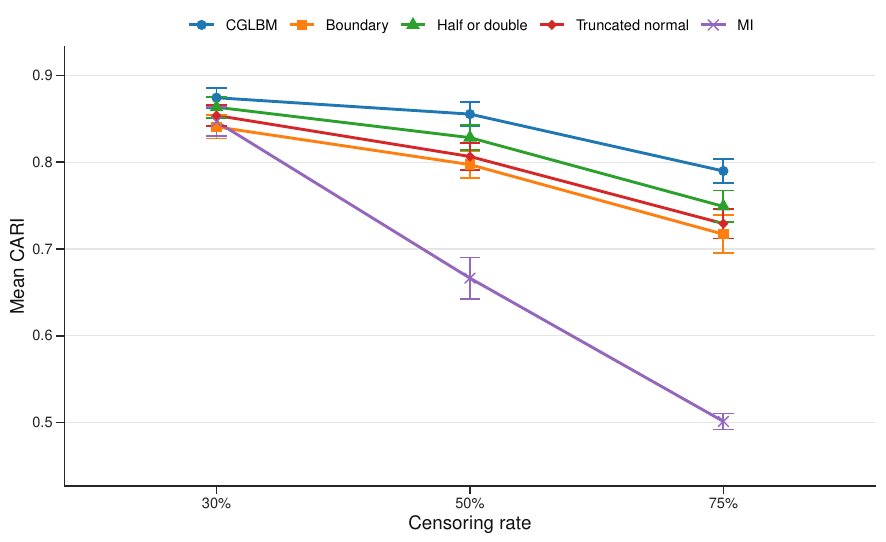}
\refstepcounter{figure}\label{figsimulationstresstest}
\par\smallskip
\begin{minipage}{0.90\textwidth}
\small\textbf{Figure \thefigure.} High-censoring stress test under \(\mathsf{SS}\). Mean CARI is shown for CGLBM and the four completion procedures at target censoring rates of 30\%, 50\%, and 75\%. The design used moderate signal, symmetric censoring, and balanced row and column class sizes. Error bars show 95\% Monte Carlo confidence intervals.
\end{minipage}
\end{figure}
\endgroup
\addtocounter{figure}{-1}

\begingroup
\renewcommand{\thetable}{B1}
\renewcommand{\theHtable}{B1}
\newcommand{\estci}[3]{\shortstack{#1\\[-1pt]{[#2, #3]}}}
\begin{table}[H]
\centering
\refstepcounter{table}\label{tabsimulationvalues}
\begin{minipage}{0.98\textwidth}
\small\textbf{Table \thetable.} Absolute CARI in Experiment~I. Values are means with 95\% Monte Carlo confidence intervals, computed separately for each signal strength.
\end{minipage}
\vspace{6pt}

\scriptsize
\setlength{\tabcolsep}{2.2pt}
\renewcommand{\arraystretch}{1.10}
\resizebox{0.99\textwidth}{!}{%
\begin{tabular}{llccccc}
\hline
Signal & Configuration & CGLBM & Boundary & Half or double & Truncated normal & MI \\
\hline
Strong & \(\mathsf{HH}\) & \estci{0.965}{0.958}{0.971} & \estci{0.945}{0.938}{0.952} & \estci{0.962}{0.955}{0.969} & \estci{0.957}{0.950}{0.964} & \estci{0.960}{0.954}{0.967} \\
Strong & \(\mathsf{HS}\) & \estci{0.965}{0.959}{0.972} & \estci{0.945}{0.938}{0.952} & \estci{0.963}{0.957}{0.969} & \estci{0.958}{0.951}{0.964} & \estci{0.961}{0.955}{0.968} \\
Strong & \(\mathsf{SH}\) & \estci{0.965}{0.958}{0.971} & \estci{0.945}{0.938}{0.952} & \estci{0.962}{0.956}{0.969} & \estci{0.957}{0.950}{0.964} & \estci{0.961}{0.954}{0.967} \\
Strong & \(\mathsf{SS}\) & \estci{0.965}{0.958}{0.972} & \estci{0.945}{0.938}{0.952} & \estci{0.963}{0.956}{0.969} & \estci{0.958}{0.951}{0.964} & \estci{0.962}{0.956}{0.968} \\

Moderate & \(\mathsf{HH}\) & \estci{0.875}{0.864}{0.885} & \estci{0.825}{0.814}{0.837} & \estci{0.862}{0.852}{0.872} & \estci{0.855}{0.845}{0.865} & \estci{0.861}{0.849}{0.872} \\
Moderate & \(\mathsf{HS}\) & \estci{0.879}{0.869}{0.889} & \estci{0.836}{0.825}{0.847} & \estci{0.870}{0.861}{0.880} & \estci{0.860}{0.851}{0.870} & \estci{0.862}{0.852}{0.873} \\
Moderate & \(\mathsf{SH}\) & \estci{0.876}{0.866}{0.886} & \estci{0.835}{0.824}{0.846} & \estci{0.868}{0.858}{0.878} & \estci{0.859}{0.849}{0.869} & \estci{0.862}{0.850}{0.874} \\
Moderate & \(\mathsf{SS}\) & \estci{0.879}{0.869}{0.889} & \estci{0.838}{0.828}{0.849} & \estci{0.871}{0.862}{0.881} & \estci{0.861}{0.851}{0.871} & \estci{0.865}{0.854}{0.876} \\

Weak & \(\mathsf{HH}\) & \estci{0.355}{0.346}{0.365} & \estci{0.319}{0.312}{0.327} & \estci{0.328}{0.320}{0.336} & \estci{0.349}{0.340}{0.358} & \estci{0.357}{0.348}{0.366} \\
Weak & \(\mathsf{HS}\) & \estci{0.375}{0.364}{0.387} & \estci{0.332}{0.324}{0.340} & \estci{0.361}{0.350}{0.371} & \estci{0.364}{0.355}{0.374} & \estci{0.363}{0.353}{0.373} \\
Weak & \(\mathsf{SH}\) & \estci{0.391}{0.383}{0.400} & \estci{0.352}{0.343}{0.361} & \estci{0.388}{0.379}{0.397} & \estci{0.379}{0.370}{0.387} & \estci{0.382}{0.374}{0.390} \\
Weak & \(\mathsf{SS}\) & \estci{0.405}{0.394}{0.415} & \estci{0.366}{0.356}{0.376} & \estci{0.404}{0.393}{0.415} & \estci{0.393}{0.382}{0.404} & \estci{0.385}{0.377}{0.394} \\
\hline
\end{tabular}%
}
\renewcommand{\arraystretch}{1}
\end{table}
\endgroup
\addtocounter{table}{-1}

\clearpage
\begingroup
\renewcommand{\thetable}{B2}
\renewcommand{\theHtable}{B2}
\refstepcounter{table}\label{tabexponeclaimsupp}
\noindent
\begin{minipage}{0.98\textwidth}
\small\textbf{Table \thetable.} Secondary recovery, block mean estimation, and interval coverage in Experiment~I. Row and Column ARI differences equal CGLBM minus the comparison procedure. Panel C reports the absolute aligned block mean RMSE for each procedure. For each completion procedure, the paired difference from CGLBM is shown in parentheses and equals the completion procedure minus CGLBM. Positive differences favor CGLBM. Panel D reports empirical coverage of $\widehat{\mu}_{k\ell}\pm1.96\,\widehat{\operatorname{se}}(\widehat{\mu}_{k\ell})$ for CGLBM under $\mathsf{SS}$ at the true $K=L=3$. Coverage was averaged over the nine blocks and three censoring directions within each latent matrix. Values in brackets are 95\% Monte Carlo confidence intervals, computed separately for each signal strength in Panels A through C and for each signal strength and censoring rate combination in Panel D.
\end{minipage}\par
\vspace{3pt}
\renewcommand{\baselinestretch}{1}\scriptsize
\setlength{\tabcolsep}{0.8pt}
\renewcommand{\arraystretch}{0.90}
\noindent
\begin{tabular*}{0.99\textwidth}{@{\extracolsep{\fill}}llcccc@{}}
\hline
Signal & Configuration & Boundary & \shortstack{Half or\\double} & \shortstack{Truncated\\normal} & MI \\
\hline
\multicolumn{6}{l}{\textit{Panel A. Row ARI difference}} \\
\hline
\multirow{4}{*}{Strong} & \(\mathsf{HH}\) & 0.007 [0.005, 0.008] & 0.001 [0.000, 0.002] & 0.002 [0.001, 0.003] & 0.001 [0.000, 0.001] \\
 & \(\mathsf{HS}\) & 0.006 [0.005, 0.008] & 0.000 [0.000, 0.001] & 0.002 [0.001, 0.003] & 0.000 [0.000, 0.001] \\
 & \(\mathsf{SH}\) & 0.007 [0.005, 0.008] & 0.000 [0.000, 0.001] & 0.002 [0.001, 0.003] & 0.000 [0.000, 0.001] \\
 & \(\mathsf{SS}\) & 0.007 [0.005, 0.008] & 0.001 [0.000, 0.001] & 0.002 [0.001, 0.003] & 0.000 [0.000, 0.001] \\
\hline
\multirow{4}{*}{Moderate} & \(\mathsf{HH}\) & 0.022 [0.017, 0.026] & 0.008 [0.003, 0.012] & 0.007 [0.005, 0.009] & 0.002 [0.000, 0.003] \\
 & \(\mathsf{HS}\) & 0.015 [0.013, 0.018] & 0.003 [0.001, 0.004] & 0.008 [0.006, 0.009] & 0.003 [0.002, 0.005] \\
 & \(\mathsf{SH}\) & 0.015 [0.012, 0.018] & 0.002 [0.001, 0.004] & 0.006 [0.005, 0.008] & 0.002 [0.000, 0.003] \\
 & \(\mathsf{SS}\) & 0.016 [0.013, 0.019] & 0.002 [0.001, 0.004] & 0.006 [0.004, 0.008] & 0.002 [0.000, 0.003] \\
\hline
\multirow{4}{*}{Weak} & \(\mathsf{HH}\) & 0.060 [0.052, 0.069] & 0.052 [0.043, 0.060] & 0.009 [0.002, 0.015] & -0.002 [-0.009, 0.005] \\
 & \(\mathsf{HS}\) & 0.050 [0.041, 0.059] & 0.018 [0.011, 0.025] & 0.012 [0.006, 0.019] & 0.007 [0.001, 0.014] \\
 & \(\mathsf{SH}\) & 0.067 [0.058, 0.075] & 0.008 [0.003, 0.014] & 0.020 [0.014, 0.025] & 0.001 [-0.003, 0.006] \\
 & \(\mathsf{SS}\) & 0.050 [0.043, 0.058] & 0.002 [-0.003, 0.006] & 0.013 [0.007, 0.019] & 0.004 [-0.001, 0.009] \\
\hline
\multicolumn{6}{l}{\textit{Panel B. Column ARI difference}} \\
\hline
\multirow{4}{*}{Strong} & \(\mathsf{HH}\) & 0.021 [0.016, 0.026] & 0.003 [0.001, 0.005] & 0.008 [0.005, 0.011] & 0.005 [0.003, 0.008] \\
 & \(\mathsf{HS}\) & 0.022 [0.017, 0.027] & 0.002 [0.000, 0.004] & 0.008 [0.005, 0.011] & 0.005 [0.003, 0.007] \\
 & \(\mathsf{SH}\) & 0.020 [0.015, 0.025] & 0.002 [0.000, 0.004] & 0.009 [0.005, 0.012] & 0.005 [0.002, 0.007] \\
 & \(\mathsf{SS}\) & 0.021 [0.016, 0.026] & 0.003 [0.001, 0.005] & 0.008 [0.005, 0.012] & 0.004 [0.002, 0.006] \\
\hline
\multirow{4}{*}{Moderate} & \(\mathsf{HH}\) & 0.053 [0.042, 0.063] & 0.012 [0.006, 0.018] & 0.022 [0.015, 0.029] & 0.019 [0.011, 0.026] \\
 & \(\mathsf{HS}\) & 0.047 [0.038, 0.056] & 0.010 [0.005, 0.014] & 0.019 [0.014, 0.025] & 0.021 [0.015, 0.027] \\
 & \(\mathsf{SH}\) & 0.044 [0.035, 0.054] & 0.009 [0.004, 0.014] & 0.018 [0.012, 0.024] & 0.018 [0.012, 0.025] \\
 & \(\mathsf{SS}\) & 0.044 [0.035, 0.053] & 0.008 [0.004, 0.013] & 0.019 [0.014, 0.025] & 0.018 [0.012, 0.024] \\
\hline
\multirow{4}{*}{Weak} & \(\mathsf{HH}\) & 0.017 [0.010, 0.025] & 0.006 [0.000, 0.013] & 0.004 [-0.002, 0.010] & -0.004 [-0.010, 0.003] \\
 & \(\mathsf{HS}\) & 0.042 [0.034, 0.050] & 0.013 [0.007, 0.020] & 0.011 [0.004, 0.018] & 0.018 [0.011, 0.025] \\
 & \(\mathsf{SH}\) & 0.017 [0.009, 0.024] & 0.000 [-0.006, 0.005] & 0.007 [0.001, 0.014] & 0.015 [0.008, 0.022] \\
 & \(\mathsf{SS}\) & 0.031 [0.024, 0.039] & -0.001 [-0.006, 0.004] & 0.011 [0.004, 0.017] & 0.034 [0.026, 0.042] \\
\hline
\end{tabular*}
\renewcommand{\arraystretch}{1}

\clearpage
\noindent
\begin{minipage}{0.98\textwidth}
\small\textbf{Table B2 continued.}
\end{minipage}\par
\vspace{3pt}
\renewcommand{\baselinestretch}{1}\scriptsize
\setlength{\tabcolsep}{0.55pt}
\renewcommand{\arraystretch}{0.90}
\noindent
\resizebox{0.99\textwidth}{!}{%
\begin{tabular}{@{}llccccc@{}}
\hline
Signal & Configuration & CGLBM & Boundary & \shortstack{Half or\\double} & \shortstack{Truncated\\normal} & MI \\
\hline
\multicolumn{7}{l}{\textit{Panel C. Aligned block mean RMSE}} \\
\hline
\multirow{4}{*}{Strong} & \(\mathsf{HH}\) & \shortstack{0.038 [0.036, 0.040]} & \shortstack{0.139 [0.137, 0.140]\\[-1pt]{(0.101 [0.099, 0.102])}} & \shortstack{0.043 [0.041, 0.044]\\[-1pt]{(0.005 [0.004, 0.005])}} & \shortstack{0.081 [0.079, 0.083]\\[-1pt]{(0.043 [0.042, 0.045])}} & \shortstack{0.040 [0.038, 0.041]\\[-1pt]{(0.002 [0.001, 0.002])}} \\
 & \(\mathsf{HS}\) & \shortstack{0.038 [0.036, 0.039]} & \shortstack{0.139 [0.137, 0.140]\\[-1pt]{(0.101 [0.099, 0.103])}} & \shortstack{0.042 [0.041, 0.044]\\[-1pt]{(0.005 [0.004, 0.005])}} & \shortstack{0.081 [0.080, 0.083]\\[-1pt]{(0.044 [0.042, 0.045])}} & \shortstack{0.040 [0.038, 0.041]\\[-1pt]{(0.002 [0.001, 0.003])}} \\
 & \(\mathsf{SH}\) & \shortstack{0.038 [0.036, 0.040]} & \shortstack{0.139 [0.137, 0.140]\\[-1pt]{(0.101 [0.099, 0.102])}} & \shortstack{0.043 [0.041, 0.044]\\[-1pt]{(0.005 [0.004, 0.005])}} & \shortstack{0.081 [0.079, 0.083]\\[-1pt]{(0.043 [0.042, 0.045])}} & \shortstack{0.040 [0.038, 0.041]\\[-1pt]{(0.002 [0.001, 0.002])}} \\
 & \(\mathsf{SS}\) & \shortstack{0.038 [0.036, 0.039]} & \shortstack{0.139 [0.137, 0.140]\\[-1pt]{(0.101 [0.099, 0.103])}} & \shortstack{0.042 [0.041, 0.044]\\[-1pt]{(0.005 [0.004, 0.005])}} & \shortstack{0.081 [0.079, 0.083]\\[-1pt]{(0.043 [0.042, 0.045])}} & \shortstack{0.040 [0.038, 0.041]\\[-1pt]{(0.002 [0.001, 0.003])}} \\
\hline
\multirow{4}{*}{Moderate} & \(\mathsf{HH}\) & \shortstack{0.043 [0.041, 0.045]} & \shortstack{0.124 [0.122, 0.126]\\[-1pt]{(0.081 [0.079, 0.083])}} & \shortstack{0.051 [0.048, 0.055]\\[-1pt]{(0.008 [0.005, 0.012])}} & \shortstack{0.077 [0.076, 0.079]\\[-1pt]{(0.034 [0.033, 0.036])}} & \shortstack{0.046 [0.044, 0.048]\\[-1pt]{(0.003 [0.002, 0.005])}} \\
 & \(\mathsf{HS}\) & \shortstack{0.042 [0.040, 0.044]} & \shortstack{0.122 [0.120, 0.123]\\[-1pt]{(0.080 [0.078, 0.082])}} & \shortstack{0.045 [0.044, 0.047]\\[-1pt]{(0.004 [0.003, 0.004])}} & \shortstack{0.077 [0.075, 0.078]\\[-1pt]{(0.035 [0.033, 0.036])}} & \shortstack{0.046 [0.044, 0.048]\\[-1pt]{(0.004 [0.003, 0.005])}} \\
 & \(\mathsf{SH}\) & \shortstack{0.043 [0.041, 0.044]} & \shortstack{0.122 [0.120, 0.124]\\[-1pt]{(0.079 [0.077, 0.081])}} & \shortstack{0.046 [0.044, 0.048]\\[-1pt]{(0.003 [0.003, 0.004])}} & \shortstack{0.077 [0.075, 0.079]\\[-1pt]{(0.034 [0.033, 0.036])}} & \shortstack{0.046 [0.044, 0.048]\\[-1pt]{(0.003 [0.002, 0.004])}} \\
 & \(\mathsf{SS}\) & \shortstack{0.042 [0.040, 0.043]} & \shortstack{0.122 [0.120, 0.123]\\[-1pt]{(0.080 [0.078, 0.082])}} & \shortstack{0.045 [0.043, 0.047]\\[-1pt]{(0.003 [0.003, 0.004])}} & \shortstack{0.077 [0.075, 0.078]\\[-1pt]{(0.035 [0.033, 0.037])}} & \shortstack{0.046 [0.044, 0.048]\\[-1pt]{(0.004 [0.003, 0.005])}} \\
\hline
\multirow{4}{*}{Weak} & \(\mathsf{HH}\) & \shortstack{0.174 [0.163, 0.185]} & \shortstack{0.210 [0.201, 0.219]\\[-1pt]{(0.036 [0.026, 0.045])}} & \shortstack{0.278 [0.263, 0.293]\\[-1pt]{(0.104 [0.090, 0.118])}} & \shortstack{0.179 [0.169, 0.189]\\[-1pt]{(0.004 [-0.003, 0.012])}} & \shortstack{0.105 [0.100, 0.110]\\[-1pt]{(-0.069 [-0.078, -0.060])}} \\
 & \(\mathsf{HS}\) & \shortstack{0.132 [0.121, 0.143]} & \shortstack{0.166 [0.158, 0.175]\\[-1pt]{(0.035 [0.026, 0.044])}} & \shortstack{0.191 [0.174, 0.207]\\[-1pt]{(0.059 [0.046, 0.072])}} & \shortstack{0.136 [0.127, 0.146]\\[-1pt]{(0.004 [-0.003, 0.012])}} & \shortstack{0.102 [0.097, 0.108]\\[-1pt]{(-0.030 [-0.037, -0.022])}} \\
 & \(\mathsf{SH}\) & \shortstack{0.138 [0.131, 0.145]} & \shortstack{0.158 [0.150, 0.165]\\[-1pt]{(0.020 [0.012, 0.027])}} & \shortstack{0.148 [0.140, 0.156]\\[-1pt]{(0.010 [0.005, 0.015])}} & \shortstack{0.140 [0.134, 0.147]\\[-1pt]{(0.002 [-0.004, 0.008])}} & \shortstack{0.089 [0.085, 0.092]\\[-1pt]{(-0.050 [-0.056, -0.043])}} \\
 & \(\mathsf{SS}\) & \shortstack{0.096 [0.090, 0.102]} & \shortstack{0.116 [0.111, 0.120]\\[-1pt]{(0.020 [0.015, 0.024])}} & \shortstack{0.099 [0.094, 0.105]\\[-1pt]{(0.003 [0.000, 0.007])}} & \shortstack{0.100 [0.094, 0.105]\\[-1pt]{(0.004 [0.000, 0.007])}} & \shortstack{0.082 [0.079, 0.085]\\[-1pt]{(-0.014 [-0.019, -0.009])}} \\
\hline
\end{tabular}%
}
\renewcommand{\arraystretch}{1}

\vspace{8pt}
\renewcommand{\baselinestretch}{1}\scriptsize
\setlength{\tabcolsep}{5pt}
\renewcommand{\arraystretch}{1.05}
\begin{center}
\begin{tabular}{@{}lccc@{}}
\hline
\multicolumn{4}{l}{\textit{Panel D. Empirical coverage of nominal 95\% intervals for fitted block means from CGLBM under $\mathsf{SS}$}} \\
\hline
Signal & 10\% & 20\% & 30\% \\
\hline
Strong & 0.93 [0.92, 0.95] & 0.93 [0.92, 0.95] & 0.93 [0.91, 0.95] \\
Moderate & 0.91 [0.89, 0.93] & 0.91 [0.89, 0.93] & 0.91 [0.89, 0.92] \\
Weak & 0.65 [0.61, 0.68] & 0.65 [0.62, 0.68] & 0.62 [0.59, 0.65] \\
\hline
\end{tabular}
\end{center}
\renewcommand{\arraystretch}{1}
\endgroup
\addtocounter{table}{-1}

\clearpage
\begingroup
\renewcommand{\thetable}{B3}
\renewcommand{\theHtable}{B3}
\refstepcounter{table}\label{tabbaselinebyfactor}
\noindent
\begin{minipage}{0.98\textwidth}
\small\textbf{Table \thetable.} Paired CARI advantages for CGLBM against each completion procedure by design factor. Panels A through C report configuration-specific results separately for each signal strength under balanced class sizes. Panel A averages over censoring conditions. Panel B averages over censoring directions within each censoring rate, and Panel C averages over censoring rates within each censoring direction. Panel D reports each class-size structure under \(\mathsf{SS}\), moderate signal, and 20\% symmetric censoring. Positive values favor CGLBM. Values in brackets are 95\% Monte Carlo confidence intervals.
\end{minipage}\par
\vspace{3pt}
\renewcommand{\baselinestretch}{1}\scriptsize
\setlength{\tabcolsep}{1.1pt}
\renewcommand{\arraystretch}{0.78}
\noindent
\resizebox{0.99\textwidth}{!}{%
\begin{tabular}{@{}l@{\hspace{4pt}}l@{\hspace{4pt}}lcccc@{}}
\hline
Signal & Level & Configuration & Boundary & \shortstack{Half or\\double} & \shortstack{Truncated\\normal} & MI \\
\hline
\multicolumn{7}{l}{\textit{Panel A. Signal strength}} \\
\hline
\multirow{4}{*}{Strong} & \multirow{4}{*}{All conditions} & \(\mathsf{HH}\) & 0.020 [0.016, 0.023] & 0.002 [0.001, 0.004] & 0.008 [0.005, 0.010] & 0.004 [0.003, 0.006] \\
 &  & \(\mathsf{HS}\) & 0.020 [0.016, 0.024] & 0.002 [0.001, 0.004] & 0.007 [0.005, 0.010] & 0.004 [0.002, 0.005] \\
 &  & \(\mathsf{SH}\) & 0.019 [0.016, 0.023] & 0.002 [0.001, 0.004] & 0.008 [0.005, 0.010] & 0.004 [0.002, 0.005] \\
 &  & \(\mathsf{SS}\) & 0.020 [0.016, 0.023] & 0.002 [0.001, 0.004] & 0.007 [0.005, 0.010] & 0.003 [0.002, 0.005] \\
\hline
\multirow{4}{*}{Moderate} & \multirow{4}{*}{All conditions} & \(\mathsf{HH}\) & 0.049 [0.041, 0.057] & 0.012 [0.007, 0.018] & 0.020 [0.015, 0.025] & 0.014 [0.009, 0.019] \\
 &  & \(\mathsf{HS}\) & 0.043 [0.036, 0.049] & 0.008 [0.005, 0.011] & 0.018 [0.014, 0.023] & 0.017 [0.012, 0.021] \\
 &  & \(\mathsf{SH}\) & 0.040 [0.033, 0.048] & 0.008 [0.004, 0.011] & 0.017 [0.012, 0.021] & 0.014 [0.009, 0.018] \\
 &  & \(\mathsf{SS}\) & 0.041 [0.034, 0.047] & 0.007 [0.004, 0.011] & 0.018 [0.014, 0.022] & 0.014 [0.009, 0.018] \\
\hline
\multirow{4}{*}{Weak} & \multirow{4}{*}{All conditions} & \(\mathsf{HH}\) & 0.036 [0.031, 0.041] & 0.027 [0.021, 0.033] & 0.006 [0.002, 0.011] & -0.002 [-0.006, 0.003] \\
 &  & \(\mathsf{HS}\) & 0.043 [0.037, 0.050] & 0.015 [0.010, 0.020] & 0.011 [0.006, 0.016] & 0.013 [0.008, 0.017] \\
 &  & \(\mathsf{SH}\) & 0.040 [0.033, 0.046] & 0.004 [0.000, 0.007] & 0.013 [0.008, 0.017] & 0.009 [0.005, 0.014] \\
 &  & \(\mathsf{SS}\) & 0.039 [0.033, 0.045] & 0.000 [-0.003, 0.004] & 0.011 [0.007, 0.016] & 0.019 [0.014, 0.024] \\
\hline
\multicolumn{7}{l}{\textit{Panel B. Censoring rate}} \\
\hline
\multirow{12}{*}{Strong} & \multirow{4}{*}{10\%} & \(\mathsf{HH}\) & 0.007 [0.004, 0.011] & 0.000 [-0.002, 0.002] & 0.001 [-0.001, 0.003] & 0.001 [-0.001, 0.003] \\
 &  & \(\mathsf{HS}\) & 0.008 [0.005, 0.012] & 0.000 [-0.002, 0.001] & 0.002 [0.000, 0.004] & 0.001 [0.000, 0.003] \\
 &  & \(\mathsf{SH}\) & 0.007 [0.004, 0.011] & 0.000 [-0.002, 0.001] & 0.002 [0.000, 0.004] & 0.002 [0.000, 0.003] \\
 &  & \(\mathsf{SS}\) & 0.007 [0.004, 0.011] & 0.000 [-0.002, 0.001] & 0.002 [0.000, 0.004] & 0.000 [-0.002, 0.001] \\
\cline{2-7}
 & \multirow{4}{*}{20\%} & \(\mathsf{HH}\) & 0.016 [0.011, 0.021] & 0.003 [0.001, 0.005] & 0.008 [0.005, 0.011] & 0.004 [0.002, 0.007] \\
 &  & \(\mathsf{HS}\) & 0.016 [0.012, 0.020] & 0.002 [0.000, 0.004] & 0.007 [0.004, 0.010] & 0.004 [0.001, 0.006] \\
 &  & \(\mathsf{SH}\) & 0.016 [0.012, 0.020] & 0.002 [0.000, 0.003] & 0.008 [0.005, 0.011] & 0.002 [0.000, 0.005] \\
 &  & \(\mathsf{SS}\) & 0.017 [0.012, 0.021] & 0.002 [0.000, 0.004] & 0.007 [0.004, 0.011] & 0.003 [0.001, 0.005] \\
\cline{2-7}
 & \multirow{4}{*}{30\%} & \(\mathsf{HH}\) & 0.036 [0.030, 0.042] & 0.005 [0.002, 0.008] & 0.014 [0.010, 0.017] & 0.007 [0.004, 0.010] \\
 &  & \(\mathsf{HS}\) & 0.036 [0.031, 0.042] & 0.005 [0.002, 0.007] & 0.013 [0.009, 0.017] & 0.007 [0.003, 0.010] \\
 &  & \(\mathsf{SH}\) & 0.035 [0.029, 0.040] & 0.005 [0.002, 0.007] & 0.013 [0.009, 0.016] & 0.007 [0.003, 0.010] \\
 &  & \(\mathsf{SS}\) & 0.035 [0.030, 0.041] & 0.005 [0.002, 0.008] & 0.013 [0.009, 0.017] & 0.006 [0.003, 0.010] \\
\hline
\multirow{12}{*}{Moderate} & \multirow{4}{*}{10\%} & \(\mathsf{HH}\) & 0.015 [0.007, 0.022] & 0.005 [0.000, 0.010] & 0.006 [0.000, 0.011] & 0.004 [-0.001, 0.008] \\
 &  & \(\mathsf{HS}\) & 0.013 [0.006, 0.020] & 0.005 [0.002, 0.007] & 0.004 [-0.001, 0.008] & 0.002 [-0.001, 0.005] \\
 &  & \(\mathsf{SH}\) & 0.016 [0.008, 0.024] & 0.005 [0.001, 0.008] & 0.006 [0.001, 0.011] & 0.003 [0.000, 0.006] \\
 &  & \(\mathsf{SS}\) & 0.012 [0.004, 0.019] & 0.004 [0.002, 0.007] & 0.005 [0.001, 0.009] & 0.003 [0.000, 0.006] \\
\cline{2-7}
 & \multirow{4}{*}{20\%} & \(\mathsf{HH}\) & 0.040 [0.030, 0.050] & 0.009 [0.001, 0.016] & 0.018 [0.010, 0.025] & 0.008 [0.003, 0.014] \\
 &  & \(\mathsf{HS}\) & 0.037 [0.030, 0.044] & 0.004 [0.000, 0.008] & 0.014 [0.008, 0.020] & 0.012 [0.007, 0.016] \\
 &  & \(\mathsf{SH}\) & 0.036 [0.028, 0.045] & 0.003 [-0.001, 0.008] & 0.013 [0.006, 0.021] & 0.007 [0.001, 0.013] \\
 &  & \(\mathsf{SS}\) & 0.035 [0.028, 0.043] & 0.004 [0.000, 0.008] & 0.017 [0.011, 0.022] & 0.008 [0.003, 0.013] \\
\cline{2-7}
 & \multirow{4}{*}{30\%} & \(\mathsf{HH}\) & 0.093 [0.078, 0.107] & 0.024 [0.012, 0.035] & 0.036 [0.028, 0.044] & 0.030 [0.020, 0.040] \\
 &  & \(\mathsf{HS}\) & 0.078 [0.067, 0.089] & 0.016 [0.010, 0.022] & 0.037 [0.030, 0.045] & 0.036 [0.027, 0.045] \\
 &  & \(\mathsf{SH}\) & 0.069 [0.058, 0.080] & 0.015 [0.009, 0.021] & 0.031 [0.023, 0.038] & 0.032 [0.022, 0.042] \\
 &  & \(\mathsf{SS}\) & 0.075 [0.064, 0.086] & 0.014 [0.008, 0.019] & 0.031 [0.025, 0.037] & 0.030 [0.020, 0.040] \\
\hline
\multirow{12}{*}{Weak} & \multirow{4}{*}{10\%} & \(\mathsf{HH}\) & 0.012 [0.004, 0.021] & 0.014 [0.006, 0.022] & 0.001 [-0.006, 0.008] & -0.013 [-0.020, -0.006] \\
 &  & \(\mathsf{HS}\) & 0.017 [0.009, 0.026] & 0.006 [-0.001, 0.014] & -0.001 [-0.007, 0.006] & 0.000 [-0.006, 0.007] \\
 &  & \(\mathsf{SH}\) & 0.010 [0.002, 0.017] & 0.001 [-0.004, 0.006] & 0.000 [-0.006, 0.005] & -0.001 [-0.006, 0.004] \\
 &  & \(\mathsf{SS}\) & 0.009 [0.002, 0.016] & -0.002 [-0.007, 0.003] & -0.001 [-0.006, 0.004] & 0.007 [0.002, 0.013] \\
\cline{2-7}
 & \multirow{4}{*}{20\%} & \(\mathsf{HH}\) & 0.035 [0.027, 0.044] & 0.018 [0.009, 0.026] & 0.000 [-0.007, 0.007] & -0.003 [-0.010, 0.003] \\
 &  & \(\mathsf{HS}\) & 0.041 [0.032, 0.051] & 0.012 [0.005, 0.020] & 0.010 [0.002, 0.018] & 0.012 [0.005, 0.019] \\
 &  & \(\mathsf{SH}\) & 0.040 [0.032, 0.047] & 0.002 [-0.003, 0.007] & 0.014 [0.008, 0.020] & 0.010 [0.004, 0.017] \\
 &  & \(\mathsf{SS}\) & 0.037 [0.028, 0.045] & -0.002 [-0.006, 0.003] & 0.010 [0.003, 0.017] & 0.019 [0.011, 0.026] \\
\cline{2-7}
 & \multirow{4}{*}{30\%} & \(\mathsf{HH}\) & 0.060 [0.051, 0.069] & 0.049 [0.038, 0.060] & 0.017 [0.009, 0.025] & 0.011 [0.004, 0.018] \\
 &  & \(\mathsf{HS}\) & 0.072 [0.063, 0.081] & 0.026 [0.017, 0.036] & 0.024 [0.015, 0.032] & 0.026 [0.019, 0.034] \\
 &  & \(\mathsf{SH}\) & 0.069 [0.061, 0.078] & 0.008 [0.001, 0.015] & 0.025 [0.019, 0.032] & 0.019 [0.012, 0.025] \\
 &  & \(\mathsf{SS}\) & 0.071 [0.062, 0.079] & 0.005 [-0.001, 0.011] & 0.026 [0.018, 0.033] & 0.032 [0.024, 0.039] \\
\hline
\end{tabular}%
}

\newpage
\noindent
\begin{minipage}{0.98\textwidth}
\small\textbf{Table B3 continued.}
\end{minipage}\par
\vspace{3pt}
\renewcommand{\baselinestretch}{1}\scriptsize
\setlength{\tabcolsep}{1.1pt}
\renewcommand{\arraystretch}{0.80}
\noindent
\resizebox{0.99\textwidth}{!}{%
\begin{tabular}{@{}l@{\hspace{4pt}}l@{\hspace{4pt}}lcccc@{}}
\hline
Signal & Level & Configuration & Boundary & \shortstack{Half or\\double} & \shortstack{Truncated\\normal} & MI \\
\hline
\multicolumn{7}{l}{\textit{Panel C. Censoring direction}} \\
\hline
\multirow{12}{*}{Strong} & \multirow{4}{*}{Right heavy} & \(\mathsf{HH}\) & 0.022 [0.016, 0.028] & 0.003 [0.000, 0.005] & 0.008 [0.005, 0.012] & 0.007 [0.004, 0.010] \\
 &  & \(\mathsf{HS}\) & 0.022 [0.016, 0.028] & 0.002 [0.000, 0.005] & 0.007 [0.003, 0.010] & 0.006 [0.003, 0.008] \\
 &  & \(\mathsf{SH}\) & 0.020 [0.015, 0.026] & 0.001 [-0.001, 0.003] & 0.007 [0.004, 0.011] & 0.006 [0.003, 0.008] \\
 &  & \(\mathsf{SS}\) & 0.022 [0.016, 0.027] & 0.003 [0.000, 0.006] & 0.007 [0.003, 0.011] & 0.005 [0.002, 0.008] \\
\cline{2-7}
 & \multirow{4}{*}{Symmetric} & \(\mathsf{HH}\) & 0.013 [0.008, 0.017] & 0.002 [-0.001, 0.005] & 0.006 [0.003, 0.009] & 0.001 [-0.001, 0.004] \\
 &  & \(\mathsf{HS}\) & 0.014 [0.009, 0.018] & 0.001 [-0.001, 0.003] & 0.007 [0.004, 0.010] & 0.003 [0.001, 0.005] \\
 &  & \(\mathsf{SH}\) & 0.013 [0.009, 0.017] & 0.003 [0.000, 0.005] & 0.007 [0.004, 0.009] & 0.003 [0.001, 0.006] \\
 &  & \(\mathsf{SS}\) & 0.014 [0.010, 0.018] & 0.002 [-0.001, 0.004] & 0.007 [0.004, 0.010] & 0.001 [-0.001, 0.004] \\
\cline{2-7}
 & \multirow{4}{*}{Left heavy} & \(\mathsf{HH}\) & 0.025 [0.019, 0.030] & 0.003 [0.000, 0.005] & 0.008 [0.005, 0.012] & 0.004 [0.001, 0.007] \\
 &  & \(\mathsf{HS}\) & 0.024 [0.019, 0.030] & 0.003 [0.001, 0.005] & 0.008 [0.005, 0.011] & 0.003 [0.000, 0.006] \\
 &  & \(\mathsf{SH}\) & 0.025 [0.019, 0.030] & 0.003 [0.000, 0.005] & 0.009 [0.005, 0.012] & 0.002 [-0.001, 0.005] \\
 &  & \(\mathsf{SS}\) & 0.024 [0.018, 0.030] & 0.002 [0.000, 0.005] & 0.008 [0.004, 0.012] & 0.003 [0.000, 0.005] \\
\hline
\multirow{12}{*}{Moderate} & \multirow{4}{*}{Right heavy} & \(\mathsf{HH}\) & 0.063 [0.051, 0.074] & 0.014 [0.005, 0.023] & 0.025 [0.018, 0.031] & 0.015 [0.010, 0.021] \\
 &  & \(\mathsf{HS}\) & 0.054 [0.045, 0.064] & 0.008 [0.003, 0.012] & 0.022 [0.016, 0.028] & 0.017 [0.011, 0.023] \\
 &  & \(\mathsf{SH}\) & 0.053 [0.043, 0.063] & 0.007 [0.002, 0.011] & 0.015 [0.009, 0.021] & 0.016 [0.009, 0.022] \\
 &  & \(\mathsf{SS}\) & 0.056 [0.046, 0.066] & 0.009 [0.004, 0.013] & 0.021 [0.015, 0.028] & 0.015 [0.009, 0.021] \\
\cline{2-7}
 & \multirow{4}{*}{Symmetric} & \(\mathsf{HH}\) & 0.028 [0.016, 0.039] & 0.009 [0.002, 0.015] & 0.014 [0.006, 0.021] & 0.014 [0.007, 0.020] \\
 &  & \(\mathsf{HS}\) & 0.024 [0.016, 0.032] & 0.008 [0.004, 0.012] & 0.014 [0.008, 0.020] & 0.015 [0.010, 0.020] \\
 &  & \(\mathsf{SH}\) & 0.022 [0.012, 0.031] & 0.006 [0.001, 0.012] & 0.013 [0.006, 0.020] & 0.009 [0.003, 0.014] \\
 &  & \(\mathsf{SS}\) & 0.022 [0.013, 0.030] & 0.006 [0.002, 0.010] & 0.013 [0.007, 0.019] & 0.013 [0.007, 0.019] \\
\cline{2-7}
 & \multirow{4}{*}{Left heavy} & \(\mathsf{HH}\) & 0.057 [0.045, 0.070] & 0.015 [0.005, 0.024] & 0.021 [0.014, 0.028] & 0.013 [0.006, 0.020] \\
 &  & \(\mathsf{HS}\) & 0.050 [0.039, 0.061] & 0.009 [0.004, 0.014] & 0.019 [0.013, 0.026] & 0.018 [0.011, 0.025] \\
 &  & \(\mathsf{SH}\) & 0.047 [0.035, 0.058] & 0.010 [0.005, 0.015] & 0.022 [0.015, 0.028] & 0.017 [0.010, 0.024] \\
 &  & \(\mathsf{SS}\) & 0.044 [0.033, 0.055] & 0.007 [0.002, 0.012] & 0.018 [0.011, 0.026] & 0.012 [0.007, 0.018] \\
\hline
\multirow{12}{*}{Weak} & \multirow{4}{*}{Right heavy} & \(\mathsf{HH}\) & 0.050 [0.041, 0.059] & 0.031 [0.022, 0.040] & 0.014 [0.006, 0.021] & 0.002 [-0.004, 0.009] \\
 &  & \(\mathsf{HS}\) & 0.055 [0.045, 0.065] & 0.021 [0.012, 0.029] & 0.019 [0.012, 0.027] & 0.013 [0.006, 0.020] \\
 &  & \(\mathsf{SH}\) & 0.057 [0.048, 0.065] & 0.004 [-0.002, 0.011] & 0.022 [0.015, 0.029] & 0.010 [0.004, 0.016] \\
 &  & \(\mathsf{SS}\) & 0.059 [0.051, 0.068] & 0.003 [-0.002, 0.008] & 0.022 [0.015, 0.029] & 0.016 [0.009, 0.023] \\
\cline{2-7}
 & \multirow{4}{*}{Symmetric} & \(\mathsf{HH}\) & 0.027 [0.019, 0.035] & 0.024 [0.016, 0.033] & 0.007 [-0.001, 0.014] & -0.003 [-0.009, 0.004] \\
 &  & \(\mathsf{HS}\) & 0.033 [0.025, 0.042] & 0.014 [0.006, 0.022] & 0.010 [0.003, 0.017] & 0.013 [0.005, 0.020] \\
 &  & \(\mathsf{SH}\) & 0.023 [0.015, 0.031] & 0.003 [-0.002, 0.008] & 0.009 [0.002, 0.015] & 0.008 [0.002, 0.014] \\
 &  & \(\mathsf{SS}\) & 0.014 [0.005, 0.022] & -0.002 [-0.007, 0.003] & 0.001 [-0.007, 0.008] & 0.019 [0.012, 0.026] \\
\cline{2-7}
 & \multirow{4}{*}{Left heavy} & \(\mathsf{HH}\) & 0.031 [0.022, 0.040] & 0.026 [0.017, 0.035] & -0.002 [-0.010, 0.006] & -0.005 [-0.012, 0.002] \\
 &  & \(\mathsf{HS}\) & 0.043 [0.033, 0.052] & 0.010 [0.002, 0.019] & 0.004 [-0.005, 0.012] & 0.012 [0.006, 0.019] \\
 &  & \(\mathsf{SH}\) & 0.039 [0.031, 0.047] & 0.004 [-0.002, 0.010] & 0.008 [0.002, 0.014] & 0.010 [0.004, 0.016] \\
 &  & \(\mathsf{SS}\) & 0.043 [0.035, 0.052] & 0.000 [-0.005, 0.005] & 0.012 [0.005, 0.019] & 0.022 [0.016, 0.029] \\
\hline
\multicolumn{7}{l}{\textit{Panel D. Class size structure under \(\mathsf{SS}\)}} \\
\hline
\multirow{4}{*}{Moderate} & Balanced & \(\mathsf{SS}\) & 0.022 [0.012, 0.033] & 0.003 [-0.002, 0.008] & 0.015 [0.007, 0.024] & 0.008 [-0.001, 0.016] \\
 & Row imbalance & \(\mathsf{SS}\) & 0.025 [0.014, 0.036] & 0.005 [-0.002, 0.012] & 0.016 [0.007, 0.025] & 0.006 [0.000, 0.012] \\
 & Column imbalance & \(\mathsf{SS}\) & 0.014 [0.005, 0.024] & -0.001 [-0.006, 0.003] & 0.004 [-0.003, 0.011] & 0.007 [0.000, 0.014] \\
 & Two way imbalance & \(\mathsf{SS}\) & 0.010 [-0.001, 0.021] & 0.001 [-0.006, 0.007] & 0.004 [-0.005, 0.012] & 0.016 [0.007, 0.025] \\
\hline
\end{tabular}%
}
\renewcommand{\arraystretch}{1}
\endgroup
\addtocounter{table}{-1}

\begingroup
\renewcommand{\thetable}{B4}
\renewcommand{\theHtable}{B4}
\begin{table}[H]
\centering
\refstepcounter{table}\label{tabconfigurationbreakdown}
\begin{minipage}{0.98\textwidth}
\small\textbf{Table \thetable.} Paired CARI advantages of \(\mathsf{SS}\) over the alternative membership configurations by signal strength, censoring rate, censoring direction, and class-size structure. Positive values favor \(\mathsf{SS}\). Values in brackets are 95\% Monte Carlo confidence intervals.
\end{minipage}
\vspace{4pt}
\renewcommand{\baselinestretch}{1}\footnotesize
\setlength{\tabcolsep}{1.4pt}
\renewcommand{\arraystretch}{1.00}
\begin{tabular*}{0.99\textwidth}{@{\extracolsep{\fill}}llccc@{}}
\hline
Signal & Level & \shortstack{\(\mathsf{SS}\) minus\\\(\mathsf{HH}\)} & \shortstack{\(\mathsf{SS}\) minus\\\(\mathsf{HS}\)} & \shortstack{\(\mathsf{SS}\) minus\\\(\mathsf{SH}\)} \\
\hline
\multicolumn{5}{l}{\textit{Panel A. Signal strength}} \\
\hline
Strong & All conditions & 0.000 [0.000, 0.001] & 0.000 [0.000, 0.000] & 0.000 [0.000, 0.001] \\
\hline
Moderate & All conditions & 0.004 [0.002, 0.007] & 0.000 [-0.001, 0.001] & 0.003 [0.001, 0.005] \\
\hline
Weak & All conditions & 0.049 [0.041, 0.057] & 0.029 [0.022, 0.036] & 0.013 [0.007, 0.019] \\
\hline
\multicolumn{5}{l}{\textit{Panel B. Censoring rate}} \\
\hline
\multirow{3}{*}{Strong} & 10\% & 0.001 [0.000, 0.002] & 0.000 [-0.001, 0.000] & 0.000 [0.000, 0.001] \\
 & 20\% & 0.001 [0.000, 0.001] & 0.000 [0.000, 0.001] & 0.001 [0.000, 0.002] \\
 & 30\% & 0.000 [-0.001, 0.001] & 0.000 [-0.001, 0.000] & 0.000 [-0.001, 0.001] \\
\hline
\multirow{3}{*}{Moderate} & 10\% & 0.002 [-0.001, 0.005] & 0.001 [-0.001, 0.003] & 0.000 [-0.003, 0.003] \\
 & 20\% & 0.004 [0.001, 0.008] & 0.000 [-0.002, 0.002] & 0.004 [0.001, 0.007] \\
 & 30\% & 0.006 [0.002, 0.010] & -0.002 [-0.004, 0.001] & 0.005 [0.002, 0.007] \\
\hline
\multirow{3}{*}{Weak} & 10\% & 0.049 [0.039, 0.058] & 0.024 [0.017, 0.032] & 0.016 [0.008, 0.024] \\
 & 20\% & 0.051 [0.041, 0.061] & 0.031 [0.023, 0.040] & 0.012 [0.004, 0.019] \\
 & 30\% & 0.048 [0.040, 0.056] & 0.032 [0.023, 0.040] & 0.012 [0.006, 0.019] \\
\hline
\multicolumn{5}{l}{\textit{Panel C. Censoring direction}} \\
\hline
\multirow{3}{*}{Strong} & Right heavy & 0.000 [0.000, 0.001] & 0.000 [-0.001, 0.000] & 0.001 [0.000, 0.002] \\
 & Symmetric & 0.001 [-0.001, 0.002] & 0.000 [0.000, 0.001] & 0.000 [-0.001, 0.001] \\
 & Left heavy & 0.000 [-0.001, 0.002] & -0.001 [-0.001, 0.000] & 0.000 [-0.001, 0.001] \\
\hline
\multirow{3}{*}{Moderate} & Right heavy & 0.006 [0.002, 0.009] & 0.002 [0.000, 0.004] & 0.004 [0.001, 0.008] \\
 & Symmetric & 0.002 [-0.001, 0.006] & -0.001 [-0.003, 0.001] & 0.003 [0.001, 0.006] \\
 & Left heavy & 0.004 [0.001, 0.008] & -0.001 [-0.004, 0.001] & 0.001 [-0.002, 0.004] \\
\hline
\multirow{3}{*}{Weak} & Right heavy & 0.043 [0.034, 0.053] & 0.028 [0.020, 0.036] & 0.012 [0.004, 0.020] \\
 & Symmetric & 0.048 [0.039, 0.057] & 0.028 [0.019, 0.036] & 0.013 [0.006, 0.020] \\
 & Left heavy & 0.056 [0.047, 0.066] & 0.032 [0.023, 0.040] & 0.014 [0.007, 0.021] \\
\hline
\multicolumn{5}{l}{\textit{Panel D. Class size structure}} \\
\hline
\multirow{4}{*}{Moderate} & Balanced & 0.003 [-0.003, 0.008] & -0.001 [-0.004, 0.002] & 0.006 [0.001, 0.011] \\
 & Row imbalance & 0.012 [0.004, 0.019] & 0.000 [-0.005, 0.005] & 0.008 [0.003, 0.013] \\
 & Column imbalance & 0.004 [-0.001, 0.008] & 0.000 [-0.003, 0.003] & 0.005 [-0.001, 0.010] \\
 & Two way imbalance & -0.003 [-0.010, 0.003] & -0.003 [-0.009, 0.003] & -0.002 [-0.008, 0.005] \\
\hline
\end{tabular*}
\renewcommand{\arraystretch}{1}
\end{table}
\endgroup
\addtocounter{table}{-1}

\begingroup
\renewcommand{\thetable}{B5}
\renewcommand{\theHtable}{B5}
\newcommand{\runtimeiqr}[3]{\shortstack{#1\\[-1pt]{[#2, #3]}}}
\begin{table}[H]
\centering
\refstepcounter{table}\label{tabexperimentiruntime}
\begin{minipage}{0.98\textwidth}
\small\textbf{Table \thetable.} CGLBM runtime by signal strength and membership configuration. Entries are median runtimes in seconds, with first and third quartiles in brackets. The benchmark was run single-threaded on an Apple M5 MacBook Air.
\end{minipage}
\vspace{6pt}

\small
\renewcommand{\arraystretch}{1.15}
\begin{tabular*}{0.94\textwidth}{@{\extracolsep{\fill}}lcccc@{}}
\hline
Signal strength & \(\mathsf{HH}\) & \(\mathsf{HS}\) & \(\mathsf{SH}\) & \(\mathsf{SS}\) \\
\hline
Strong
& \runtimeiqr{0.571}{0.541}{0.599}
& \runtimeiqr{0.664}{0.619}{0.743}
& \runtimeiqr{0.613}{0.596}{0.648}
& \runtimeiqr{0.819}{0.757}{0.911} \\
Moderate
& \runtimeiqr{0.623}{0.594}{0.659}
& \runtimeiqr{0.927}{0.831}{1.024}
& \runtimeiqr{0.692}{0.664}{0.736}
& \runtimeiqr{1.143}{1.011}{1.280} \\
Weak
& \runtimeiqr{0.663}{0.629}{0.699}
& \runtimeiqr{2.318}{2.093}{2.547}
& \runtimeiqr{1.596}{1.421}{2.058}
& \runtimeiqr{3.856}{3.099}{4.855} \\
\hline
\end{tabular*}
\renewcommand{\arraystretch}{1}
\end{table}
\endgroup
\addtocounter{table}{-1}

\subsection{Experiment II calibration and matching details}\label{app:experiment_two_design}

\paragraph{Signal calibration.}
Take \(\mathbf s_2=\sqrt{2/3}(-1,1)^\top\) and
\(\mathbf s_3=(-1,0,1)^\top\).
For a true pair \((K^\star,L^\star)\), let
\(\mathbf r=\mathbf s_{K^\star}\) and
\(\mathbf q=\mathbf s_{L^\star}\).
The interaction matrix \(\mathbf G\) equaled \(\boldsymbol\Gamma\) for \((3,3)\). When one dimension had two classes, the endpoint rows or columns of \(\boldsymbol\Gamma\) were selected and the resulting matrix was centered across rows and columns. The block parameters were
\[
\begin{aligned}
\mu_{k\ell}^\star
&= a_Rr_k+a_Cq_\ell+0.25\sqrt{a_Ra_C}\,G_{k\ell},\\
(\sigma_{k\ell}^\star)^2
&= \frac{\exp(0.25G_{k\ell})}
{(K^\star L^\star)^{-1}
\sum_{u=1}^{K^\star}\sum_{v=1}^{L^\star}\exp(0.25G_{uv})}.
\end{aligned}
\]
For \(P_r=\normal(\mu_r,v_r)\), define the symmetric Gaussian KL divergence by
\[
d_{\mathrm{SKL}}(P_1,P_2)
=
\frac14
\left\{
\frac{v_1}{v_2}
+
\frac{v_2}{v_1}
-2
+
(\mu_1-\mu_2)^2
\left(
\frac{1}{v_1}
+
\frac{1}{v_2}
\right)
\right\}.
\]
The minimum row and column class separations were
\[
\begin{aligned}
D_R
&=
\min_{k<k'}
\frac{1}{L^\star}
\sum_{\ell=1}^{L^\star}
d_{\mathrm{SKL}}
\left\{
\normal(\mu_{k\ell}^\star,(\sigma_{k\ell}^\star)^2),
\normal(\mu_{k'\ell}^\star,(\sigma_{k'\ell}^\star)^2)
\right\},\\
D_C
&=
\min_{\ell<\ell'}
\frac{1}{K^\star}
\sum_{k=1}^{K^\star}
d_{\mathrm{SKL}}
\left\{
\normal(\mu_{k\ell}^\star,(\sigma_{k\ell}^\star)^2),
\normal(\mu_{k\ell'}^\star,(\sigma_{k\ell'}^\star)^2)
\right\}.
\end{aligned}
\]
For the \((3,3)\) reference, the strong, moderate, and weak signal scales were
\[
(a_R,a_C)
\in
\{(0.75,0.400),(0.60,0.320),(0.35,0.185)\},
\]
respectively. The scales for \((2,3)\) and \((3,2)\) were chosen to match both \(D_R\) and \(D_C\) of the corresponding \((3,3)\) signal setting. Table~\ref{tabexperimentiicalibration} gives the calibrated scales and target separations.

\paragraph{Class sizes and matching.}
Balanced class sizes were \(60,60\) rows and \(20,20,20\) columns for \((2,3)\), \(40,40,40\) rows and \(30,30\) columns for \((3,2)\), and \(40,40,40\) rows and \(20,20,20\) columns for \((3,3)\). Row imbalance used \(70,50\) for two row classes and \(50,40,30\) for three row classes. Column imbalance used \(35,25\) for two column classes and \(25,20,15\) for three column classes. Two-way imbalance combined the corresponding row and column structures. The class-size vectors were cyclically permuted across replicates so that each signal profile occurred at each class size. Each rotated positive-censoring specification was calibrated separately.

For a fixed true pair, signal strength, class-size structure, and replicate, the same latent matrix was evaluated at all four censoring rates in the main design. Under moderate signal, balanced classes, and 20\% censoring, the same latent matrix was also evaluated under the three censoring directions. For each candidate class-number pair, these matched conditions used the same ordered pool of initial row and column partitions. Positive censoring rates used independent calibration matrices.

\begingroup
\renewcommand{\thetable}{B6}
\renewcommand{\theHtable}{B6}
\begin{table}[H]
\centering
\refstepcounter{table}\label{tabexperimentiicalibration}
\begin{minipage}{0.98\textwidth}
\small\textbf{Table \thetable.} Calibrated signal scales and target row and column separations in Experiment~II.
\end{minipage}
\vspace{6pt}

\scriptsize
\setlength{\tabcolsep}{5pt}
\renewcommand{\arraystretch}{1.10}
\begin{tabular*}{0.96\textwidth}{@{\extracolsep{\fill}}llrrrr@{}}
\hline
Signal & True \((K^\star,L^\star)\) & \(a_R\) & \(a_C\) & Target \(D_R\) & Target \(D_C\) \\
\hline
Strong & \((2,3)\) & 0.465326 & 0.414544 & 0.343441 & 0.134186 \\
Strong & \((3,2)\) & 0.763544 & 0.254494 & 0.343441 & 0.134186 \\
Strong & \((3,3)\) & 0.750000 & 0.400000 & 0.343441 & 0.134186 \\
Moderate & \((2,3)\) & 0.372355 & 0.327757 & 0.231229 & 0.097306 \\
Moderate & \((3,2)\) & 0.608657 & 0.203644 & 0.231229 & 0.097306 \\
Moderate & \((3,3)\) & 0.600000 & 0.320000 & 0.231229 & 0.097306 \\
Weak & \((2,3)\) & 0.217500 & 0.176681 & 0.099586 & 0.053693 \\
Weak & \((3,2)\) & 0.348078 & 0.117925 & 0.099586 & 0.053693 \\
Weak & \((3,3)\) & 0.350000 & 0.185000 & 0.099586 & 0.053693 \\
\hline
\end{tabular*}
\renewcommand{\arraystretch}{1}
\end{table}
\endgroup
\addtocounter{table}{-1}

\subsection{Additional results for Experiment II}\label{app:experiment_two_additional_results}

\begingroup
\renewcommand{\thefigure}{B4}
\renewcommand{\theHfigure}{B4}
\begin{figure}[H]
\centering
\includegraphics[width=0.99\textwidth]{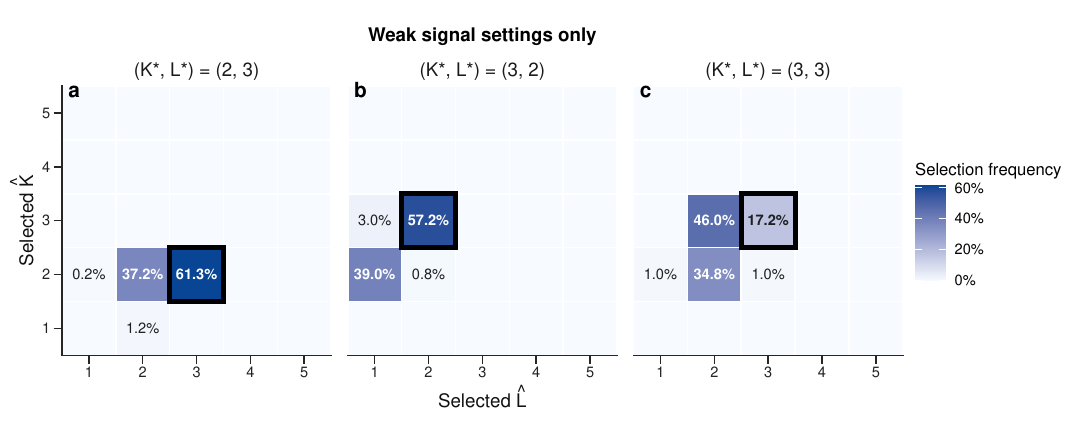}
\refstepcounter{figure}\label{figexperimentiifailures}
\par\smallskip
\begin{minipage}{0.98\textwidth}
\small\textbf{Figure \thefigure.} Weak-signal class-number selection patterns. Each panel reports the selected-pair frequencies for one true class-number pair across 400 report matrices. The black outline marks the true pair, and all panels use a common frequency scale.
\end{minipage}
\end{figure}
\endgroup
\addtocounter{figure}{-1}

\begingroup
\renewcommand{\thetable}{B7}
\renewcommand{\theHtable}{B7}
\begin{table}[H]
\centering
\refstepcounter{table}\label{tabexperimentiifittingdiagnostics}
\begin{minipage}{0.98\textwidth}
\small\textbf{Table \thetable.} Weak-signal exact-pair recovery, selection error structure, and candidate eligibility in Experiment~II.
\end{minipage}
\vspace{6pt}

\scriptsize
\renewcommand{\arraystretch}{1.12}
\textit{Panel A. Weak-signal exact-pair recovery}\par\smallskip
\begin{tabular*}{0.94\textwidth}{@{\extracolsep{\fill}}lcccc@{}}
\hline
True \((K^\star,L^\star)\) & 0\% (reference) & 10\% & 20\% & 30\% \\
\hline
\((2,3)\) & 86/100 (86\%) & 70/100 (70\%) & 55/100 (55\%) & 34/100 (34\%) \\
\((3,2)\) & 82/100 (82\%) & 66/100 (66\%) & 48/100 (48\%) & 33/100 (33\%) \\
\((3,3)\) & 38/100 (38\%) & 19/100 (19\%) & 10/100 (10\%) & 2/100 (2\%) \\
\hline
\end{tabular*}

\vspace{8pt}
\textit{Panel B. Selection errors (\(n=658\))}\par\smallskip
\begin{tabular*}{0.84\textwidth}{@{\extracolsep{\fill}}lc@{}}
\hline
Outcome & Count and percentage \\
\hline
Errors under weak signal & 657/658 (99.85\%) \\
Too few column classes only & 347/658 (52.74\%) \\
Too few row classes only & 7/658 (1.06\%) \\
Too few classes in both dimensions & 304/658 (46.20\%) \\
Any selection of too many classes & 0/5100 (0\%) \\
\hline
\end{tabular*}

\vspace{8pt}
\textit{Panel C. Candidate eligibility}\par\smallskip
\begin{tabular*}{0.84\textwidth}{@{\extracolsep{\fill}}lc@{}}
\hline
Outcome & Count and percentage \\
\hline
Excluded candidates & 2053/127500 (1.61\%) \\
Excluded true candidates & 0/5100 (0\%) \\
\hline
\end{tabular*}
\renewcommand{\arraystretch}{1}
\end{table}
\endgroup
\addtocounter{table}{-1}

\clearpage

\end{document}